\documentclass[preprint,11pt]{JHEP3} 

\JHEPspecialurl{http://jhep.sissa.it/JOURNAL/JHEP3.tar.gz}

\usepackage{epsfig,multicol,amsmath}

\usepackage{epstopdf}
\usepackage{rotating}
\usepackage{cite}

\def\beq{\begin{equation}}
\def\eeq{\end{equation}}
\def\bea{\begin{eqnarray}}
\def\eea{\end{eqnarray}}

\def\eq#1{{Eq.~(\ref{#1})}}
\def\fig#1{{Fig.~\ref{#1}}}

 \newcommand{\SP}{\langle \mid S_{H} \mid^2 \rangle}

\newcommand{\Lb}{\left(}
\newcommand{\Rb}{\right)}
\parskip=\itemsep               

\newcommand{\A}{{\cal A}}

\def\ap#1#2#3{     {\it Ann. Phys. (NY) }{\bf #1} (19#2) #3}
\def\arnps#1#2#3{  {\it Ann. Rev. Nucl. Part. Sci. }{\bf #1} (19#2) #3}
\def\npb#1#2#3{    {\it Nucl. Phys. }{\bf B#1} (19#2) #3}
\def\plb#1#2#3{    {\it Phys. Lett. }{\bf B#1} (19#2) #3}
\def\prd#1#2#3{    {\it Phys. Rev. }{\bf D#1} (19#2) #3}
\def\prep#1#2#3{   {\it Phys. Rep. }{\bf #1} (19#2) #3}
\def\prl#1#2#3{    {\it Phys. Rev. Lett. }{\bf #1} (19#2) #3}
\def\ptp#1#2#3{    {\it Prog. Theor. Phys. }{\bf #1} (19#2) #3}
\def\rmp#1#2#3{    {\it Rev. Mod. Phys. }{\bf #1} (19#2) #3}
\def\zpc#1#2#3{    {\it Z. Phys. }{\bf C#1} (19#2) #3}
\def\mpla#1#2#3{   {\it Mod. Phys. Lett. }{\bf A#1} (19#2) #3}
\def\nc#1#2#3{     {\it Nuovo Cim. }{\bf #1} (19#2) #3}
\def\yf#1#2#3{     {\it Yad. Fiz. }{\bf #1} (19#2) #3}
\def\cpc#1#2#3{    {\it Comp. Phys. CommunTable. }{\bf #1} (19#2) #3}

\title{A Soft Interaction Model at Ultra High Energies:\\
 Amplitudes, Cross Sections and Survival Probabilities}
\author{\Large  E. Gotsman\thanks{Email:
gotsman@post.tau.ac.il.}\,, E. Levin\thanks{Email:
leving@post.tau.ac.il,
levin@mail.desy.de.} \,and\, U. Maor\thanks{Email:
maor@post.tau.ac.il.}\\
Department of Particle Physics, School of Physics and Astronomy\\ 
Raymond and Beverly Sackler Faculty of Exact Science\\  
Tel Aviv University, Tel Aviv, 69978, Israel}

\bigskip

\abstract{
In this paper we present a two channel  model with the goal of 
reproducing the soft
scattering data available in the ISR-Tevatron energy range,
and extend the model
results to  LHC and Cosmic Rays energies.  A characteristic feature of 
the model is that we represent the sum of all diffractive final states at
a vertex, by a single diffractive state.
Our two main results are:
(i) The approach of the  
elastic scattering amplitude  to the black disc bound
is  very slow, reaching it at energies far higher than the GZK ankle
cutoff. 
(ii) Our predicted survival probability for Higgs exclusive central
diffractive production at the LHC is 0.7\%, 
which is considerably smaller than our previous estimate. 
The above features are compatible with a parton-like model in which the
traditional soft Pomeron is replaced by an amplitude describing the 
partonic system, which is
saturated in the soft (long distance) limit. 
}

\dedicated{PACS: 13.85.-t, 13.85.Hd, 11.55.-m, 11.55.Bq}
\preprint{TAUP -2560-07\\
\today}
\begin{document}
\section{Introduction}
\par
The GLM model is based on a two channel eikonal like solution of the
$s$-channel unitarity equation\cite{2CH}. Our present investigation is based on 
an improved model, and we calculate and discuss it's  predictions and 
implications 
at the LHC and Cosmic Rays energies.
For the correct degrees of freedom, 
the partial amplitude for scattering of state `$i$' with state `$k$' can be 
written as
\beq \label{MOD1}
A^{i',k'}_{i,k}(s,b)\,=\,i\,\delta_{i,i'} \delta_{k,k'}\,
\Lb 1\,\,-\,\,\exp\Lb - \frac{\Omega_{i,k}(s,b)}{2} \Rb \Rb.
\eeq
i and k represent sets of quantum numbers which diagonalize the
interaction matrix. The opacities 
$\Omega_{i,k}$ are arbitrary real functions of energy and impact 
parameter. 
\par
As long as $\Omega_{i,k}$  are not explicitly specified, 
our presentation is  model independent.  We 
assume that the scattering amplitude at high energies is predominantly 
imaginary.
In our original single channel
model\cite{1CH} we assumed that the opacity is determined by the exchange
of a soft Pomeron represented by a factorized fixed pole in the complex
angular momentum  (J) plane.
 This simplified assumption is not maintained in 
the present, more elaborate
two channel model, where 
 factorization of the coupling constants is relaxed.
These features are compatible with our present partonic picture, where the soft
Pomeron, is replaced by the soft distance limit, of the amplitude for  a 
saturated partonic
system \cite{SAT}. This approach is supported by eikonal-like models
\cite{DISMOD,KOTE}, which reproduce the e-p DIS data over a wide range of 
$Q^{2}$,
starting from very small virtualities. Details of our parametrization are
presented in Sec. 2. 
\par
We shall elaborate on the following issues:
\newline
1) Our original investigation\cite{2CH} neglected the double diffraction  
state. Experimental data on this channel is now available\cite{DDD} and it
makes the simplified two amplitude approximation doubtful.
The present three amplitude analysis is based on an updated data base
which includes the published $p$-$p$ and $\bar{p}$-$p$ data points of
$\sigma_{tot}$, the integrated values of $\sigma_{el}$, $\sigma_{sd}$,
$\sigma_{dd}$ and the forward elastic slope $B_{el}$ in the ISR-Tevatron
energy range. 
The forward slopes of the SD and DD final states, as well as
$\rho=\frac{Re\,a_{el}(t=0,s)}{Im\,a_{el}(t=0,s)}$,  
are predictions of the model.
\newline
2) Based on this initial investigation, our present analysis aims at 
providing  reliable
predictions of the quantities to be measured at  14 $TeV$, the LHC c.m. 
energy.
Our output also covers the broad Cosmic Ray energy range up to the GZK 
limit.
We validate the consistency of our calculations with 
unitarity by extending our output up to the Planck mass.
\newline
3) One requires a reliable formulation of soft scattering
  to calculate  
the survival probability of large rapidity gaps (LRG), initiated by
the underlying soft rescatterings of the spectator partons, in an inelastic 
diffractive (soft or hard) process\cite{Bj,GLM1,1CH}.
This calculation is of particular importance
for the assessment of the discovery potential for LHC Higgs production in an
exclusive central diffractive process. This channel,
with a clean two LRG signature, has a relatively good signal
to background ratio.
The extraction of a clear diffractive Higgs signal at the LHC, requires a 
precise
knowledge of the cross section and transverse momentum behavior
of the single diffractive production channel, which provides a significant
contribution to the background of interest. 
We shall investigate the SD channel in some detail.
\newline
4) Some of the fundamental consequences of
s-channel unitarity in the high energy limit are not clear as yet.
We wish to assess the rate that the elastic scattering amplitude reaches 
the unitarity black disc bound in $b$-space.
This is, obviously, coupled to the behavior of 
the corresponding diffractive amplitudes.
A related piece of information, which is still unknown,
is the rate at which the proton black core  
expands with energy in $b$-space.
At present, different models provide drastically 
different assessment of this phenomenon. 
Our advantage of having a specific model enables us to study in detail
the behavior of the amplitudes as functions of energy and impact 
parameter, and, consequently, we are able to provide a numerical 
description 
of the process. 
Our approach differs from alternative treatments which are based on   
general assessments which are not accompanied by a detailed analysis 
of the soft scattering data. See, for example Ref.\cite{S2FS}. 
\par
The organization of this paper is as follows:
In Sec.2 we briefly summarize 
the general  properties of the eikonal approach and formulate our  model.
Following, we review the extension from GLM 
single channel to two channels. 
In Sec.3 we present the three models we have considered and 
the output of our calculations in the ISR-Tevatron energy range. 
Sec.4 is devoted to a presentation of 
our predictions for LHC and Cosmic Ray energies. 
In Sec.5 we discuss survival probability calculations in the GLM models   
applied to Higgs production at LHC.  
Sec.6 is devoted to a detailed analysis of the onset of unitarity effects 
at exceedingly high energies, and the approach to the black disc bound in 
$b$-space. Our conclusions are presented in Sec.7.

\section{The GLM Model}
\subsection{ Single channel model}
\par
The main assumption of the single channel GLM model is 
that hadrons are the 
correct degrees of freedom at high energy, diagonalizing the scattering matrix.
This  model\cite{1CH} fits 
$\sigma_{tot}$, $\sigma_{el}$ and $B_{el}$ well,
but fails to reproduce the inelastic diffractive final states.
This is evident in the relatively better measured SD channel, in which the
calculated normalization and energy dependence of $\sigma_{sd}$ fail to
agree with the experimental data\cite{1CH}. This is not
surprising as the input assumption of this class of models, that
$\frac{\sigma_{sd}}{\sigma_{el}}$ is negligibly small,
is not compatible with the data\cite{Dino}.
\DOUBLEFIGURE[ht]{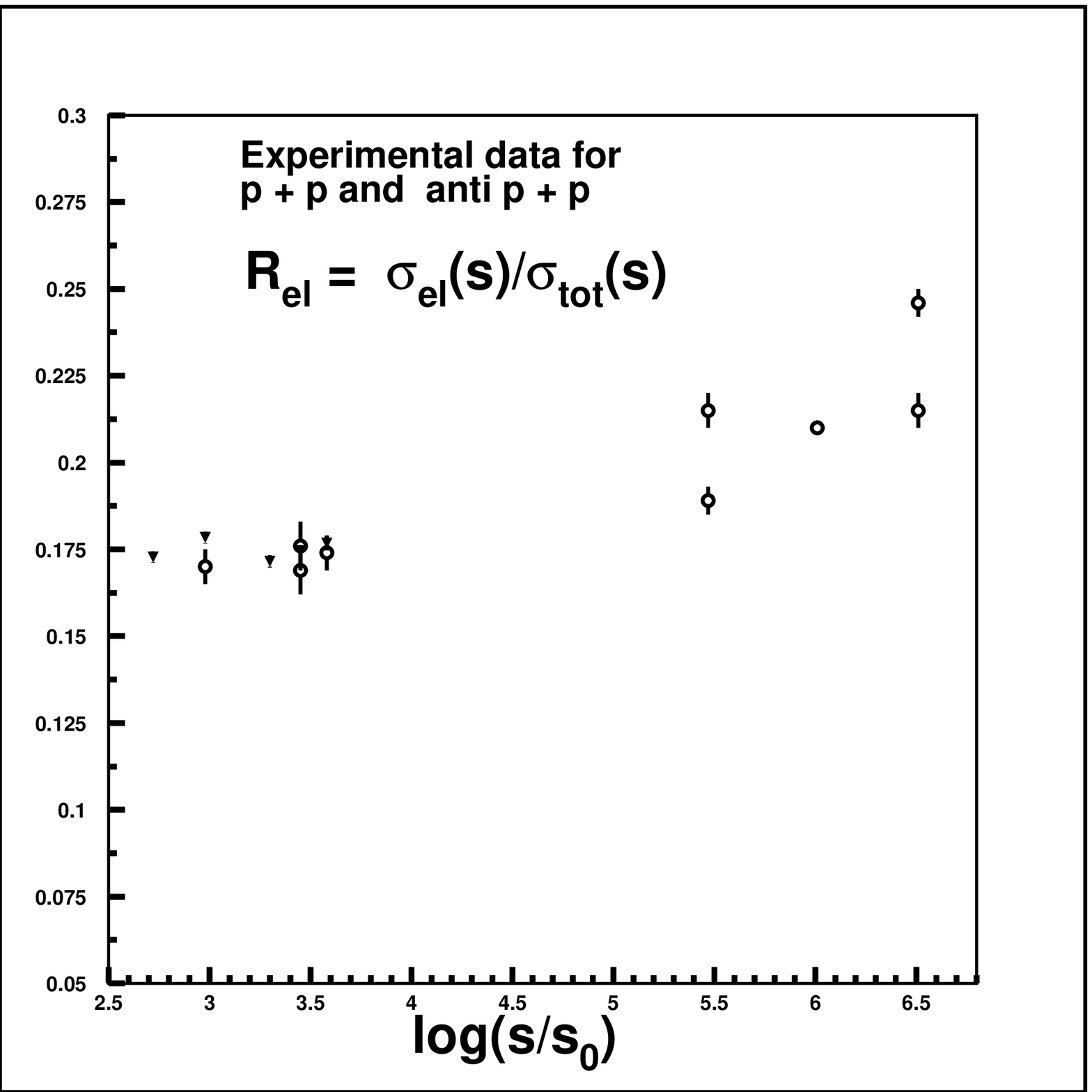,width=90mm,height=80mm}
{ 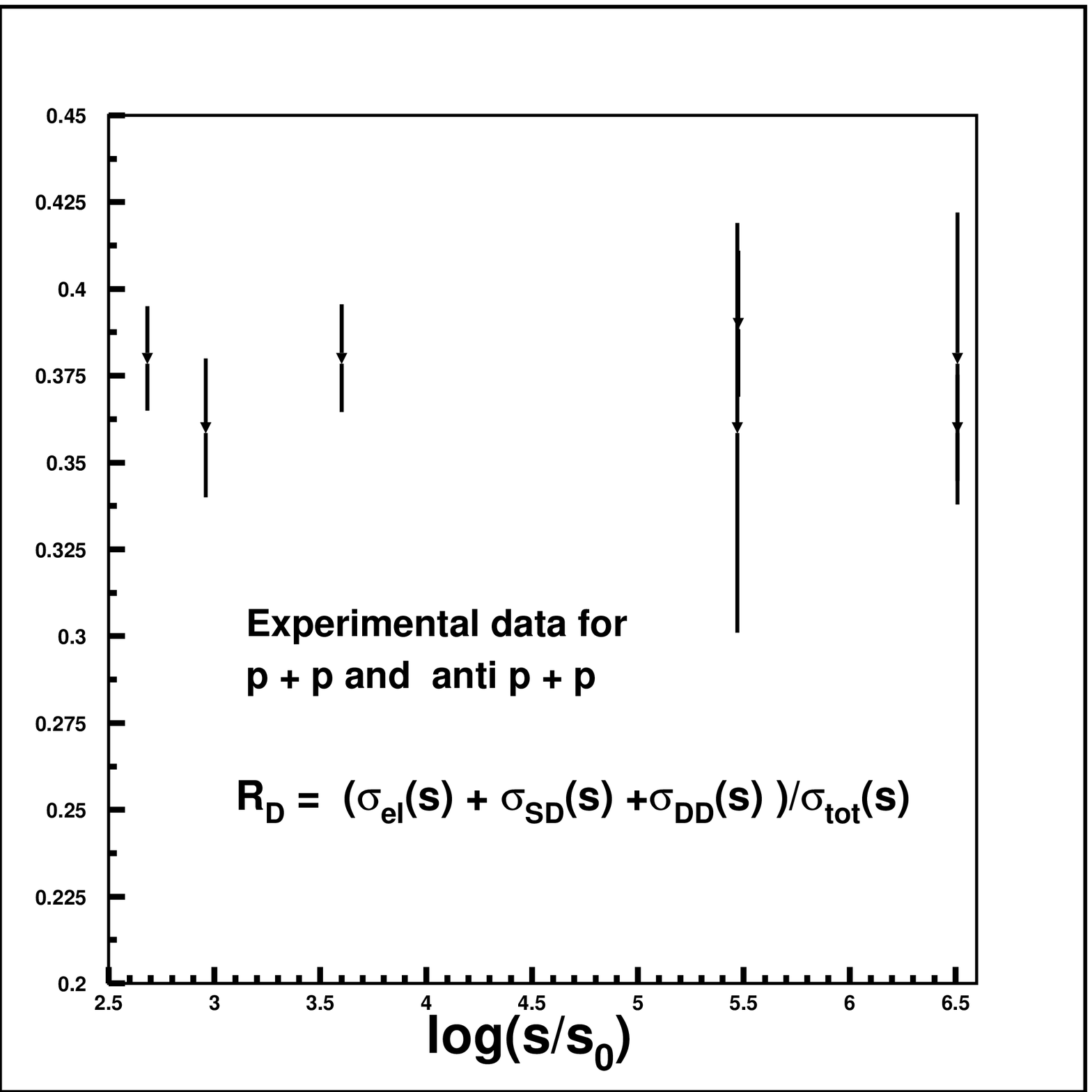,width=90mm,height=80mm}
{Energy dependence of $R_{el}$.
\label{reltot}}
{Energy dependence of $R_{D}$.
\label{rdtot}}
\par
It is instructive to see the ISR-Tevatron
experimental data presented in \fig{reltot} and \fig{rdtot}.
\fig{reltot} shows the power like increase  of the ratio
$R_{el}=\frac{\sigma_{el}}{\sigma_{tot}}$ with $s$  ,
which is compatible with  our parametrization 
(soft Pomeron like) , and can be reproduced in a single channel eikonal
model in the ISR-LHC energy range\cite{Zur}. 
Defining $\sigma_{diff}=\sigma_{sd}+\sigma_{dd}$, 
the ratio
$R_{D}=\frac{\sigma_{el}+\sigma_{diff}}{\sigma_{tot}}$, which is 
shown in \fig{rdtot},
behaves approximately as a constant of about 0.37-0.38. This
is incompatible with the assumed exchange of an unscreened
soft Pomeron, where, $R_{el}$ and $R_{D}$ are expected to
have approximately the same energy dependence. 
In a single channel eikonal model $\frac{\sigma_{diff}}{\sigma_{el}}$ 
is a small parameter.
i.e., diffractive scattering is treated as a perturbative effect.
As such, this ratio tends to  a small constant at low energies, and
approaches zero in the high energy limit. 
This behavior is in contradiction with the
experimental data shown in \fig{rdtot}.
\par
In the following we show
that the deficiencies of the single channel eikonal model are eliminated in
a more elaborate two channel model in which diffractive,
alongside elastic, scatterings are included
in the rescattering chain.

\subsection{Two channel model}
\par
The GLM two channel model has been described in our previous publications
(see Refs.\cite{2CH,SP2CH,GKLMP,heralhc} and references therein).
In this formalism, diffractively produced hadrons at a given vertex are
considered as a single hadronic state
described by the wave function $\Psi_D$, which is orthonormal
to the wave function $\Psi_h$ of the incoming hadron (proton in the case of
interest), $<\Psi_h|\Psi_D>=0 $.
We introduce two wave functions $\Psi_1$ and $\Psi_2$ that diagonalize the
2x2 interaction matrix ${\bf T}$
\beq \label{2CHM}
A_{i,k}^{i',k'}=<\Psi_{i}\,\Psi_{k}|\mathbf{T}|\Psi_{i'}\,\Psi_{k'}>=
A_{i,k}\,\delta_{i,i'}\,\delta_{k,k'},
\eeq
In this representation the observed states are written in the form
\beq \label{2CHM31}
\Psi_h=\alpha\,\Psi_1+\beta\,\Psi_2\,,
\eeq
\beq \label{2CHM32}
\Psi_D=-\beta\,\Psi_1+\alpha \,\Psi_2\,,
\eeq
where, $\alpha^2+\beta^2=1$.
\par
Using \eq{2CHM} we can rewrite the unitarity equations in the form
\beq \label{UNIT}
Im\,A_{i,k}\left(s,b\right)=|A_{i,k}\left(s,b\right)|^2
+G^{in}_{i,k}(s,b),
\eeq
where $G^{in}_{i,k}$ is the contribution of all non diffractive inelastic 
processes. 
i.e., it is the summed probability for these final states to be 
produced in the scattering of particle $i$ off particle $k$.
The simple solution to \eq{UNIT} has the same structure as in the single channel 
formalism,  
\beq \label{2CHM1}
A_{i,k}(s,b)=i \Lb 1 -\exp\Lb - \frac{\Omega_{i,k}(s,b)}{2}\Rb\Rb, 
\eeq
\beq \label{2CHM2}
G^{in}_{i,k}(s,b)=1-\exp\Lb - \Omega_{i,k}(s,b)\Rb.
\eeq
From \eq{2CHM2} we deduce  the probability that the initial projectiles
$(i,k)$ reach the final state interaction unchanged, regardless of the initial
state rescatterings, is given by
$P^S_{i,k}=\exp \Lb - \Omega_{i,k}(s,b) \Rb$.
\par
Our presentation, thus far, is model independent.
Model dependent parameters are introduced so as to obtain explicit
expressions for the opacities $\Omega_{i,k}(s,b)$. 
As stated in the Introduction, we now replace the simple Pomeron J-pole 
dynamic input, by an updated interpretation, in which the Pomeron is 
replaced by the description 
 of a partonic system, which goes through a
process of saturation when approaching the soft, large distance 
limit\cite{SAT}.
This point of view is compatible with eikonal type models which reproduce
DIS e-p scattering data  over the whole  $Q^{2}$ range,  i.e. from
low to high photon virtualities\cite{DISMOD,KOTE}.
\par
Even though we now suggest a more elaborate partonic description for the
dynamics of interest, we still maintain a simple form  
for the opacities $\Omega_{i,k}$,
\beq \label{omega}
\Omega_{i,k}\Lb s,b \Rb = \nu_{i,k}\Lb s \Rb \Gamma\Lb s,b \Rb. 
\eeq
In the above 
\beq
\nu_{i,k}\Lb s \Rb = \sigma^0_{i,k}\,\Lb \frac{s}{s_0}\Rb^{\Delta},
\eeq
and
\beq
\Gamma_{i,k} \Lb s,b \Rb = \,\,\frac{1}{\pi R^2_{i,k}\Lb s \Rb}
\exp \Lb - \frac{b^2}{ R^2_{i,k}\Lb s \Rb}\Rb.
\eeq
The energy term $\nu_{i,k}$ is a power in $s$ reflecting a long standing 
observation\cite{DL} that the cross section of DIS 
collisions behaves, at least approximately, as a power in energy.
The  b-profile $\Gamma_{i,k} \Lb s,b \Rb$ are assumed to be Gaussians.
Since the t-space transform of a Gaussian in impact
parameter is proportional to $\exp \Lb -\frac{R^2_{i,k}}{4}|t| \Rb$,
our parametrization provides a very
good reproduction of the small $|t|$ elastic forward cone data,
covering more than 95\% of the elastic scattering events.  
A general property of a b-Gaussian profile is that the  small $t$ region 
transforms to high $b$, and high $t$ to small $b$. 
The structure of our opacities, as presented in \eq{omega}, 
has the advantage that it is general enough to serve as a parametrization 
of wide class of dynamical models.  
It is applicable  both to a conventional soft Pomeron exchange, as
well as a hard Pomeron process such as
$\gamma + p \rightarrow J/\Psi + p$\cite{PSISL}.
\par
Even though we now suggest a more elaborate partonic description for the 
dynamics of interest, we still choose for our radius a Regge-like 
expression
\beq \label{radius}
R^2_{i,k}\Lb s \Rb = R^2_{0,i}+ R^2_{0,k}+4\alpha'\ln(s/s_0),
= R^2_{0;i,k} + 4 \alpha^{'} ln(\frac{s}{s_{0}})
\eeq
in which $R^2_{0,j}$ and $\alpha'$ are fitted parameters. 
Indeed, the fitted values of $\alpha'$ 
corresponding to Models B(1) and B(2), to be discussed in the next Section, 
differ from the values typical to Regge phenomenology.
Note that $R^2_{0,2}=0$.
\par

In general, we have to consider four
possible re-scattering processes.  For the case of
$p$-$p$ (or $\bar p$-$p$) the two quasi-elastic amplitudes are equal
$a_{1,2}=a_{2,1}$, and we  thus have three rescattering amplitudes:
elastic, SD and DD.
These amplitudes are presented
in the two channel formalism in
the following form\cite{2CH,SP2CH,heralhc}
\beq \label{EL}
a_{el}(s,b)=
i\{\alpha^4A_{1,1}+2\alpha^2\beta^2A_{1,2}+\beta^4\A_{2,2}\},
\eeq
\beq \label{SD}
a_{sd}(s,b)=
i\alpha\beta\{-\alpha^2A_{1,1}+(\alpha^2-\beta^2)A_{1,2}+\beta^2A_{2,2}\},
\eeq
\beq \label{DD}
a_{dd}=
i\alpha^2\beta^2\{A_{1,1}-2A_{1,2}+A_{2,2}\}.
\eeq
It should be stressed that in this approach diffraction dissociation  
appears as an outcome of the elastic scattering 
of $\Psi_{1}$ and $\Psi_{2}$, the 
correct degrees of freedom of our model.
\par
The corresponding cross sections are given by
\beq
\sigma_{tot}(s)=2\int d^2 b \,Im\,a_{el}\Lb s,b\Rb,
\eeq
\beq
\sigma_{el}(s)=\int d^2 b \,|a_{el}\Lb s,b\Rb|^2,
\eeq
\beq
\sigma_{sd}(s)=\int d^2 b \,|a_{sd}\Lb s,b\Rb|^2,
\eeq
\beq
\sigma_{dd}(s)=\int d^2 b \,|a_{dd}\Lb s,b\Rb|^2.
\eeq
\par
In a simplified version we considered a two amplitude model,
where $a_{dd}$ is assumed to be small enough to be neglected implying 
that 
$A_{2,2}=2A_{1,2}-A_{1,1}$.
In this model\cite{2CH}
only elastic and SD rescatterings are included in the eikonal screening 
correction.
In our past publications we referred to the GLM eikonal models
according to the number of the rescattering channels considered, i.e.
elastic\cite{1CH},
elastic+SD\cite{2CH} and elastic+SD+DD\cite{SP2CH}. In retrospect, we
consider it more appropriate to define these models according to the
dimensionality of their amplitude base. We, therefore, call the 2x2 configuration 
a two channel model, making the distinction between its two and three amplitude
representations.

\section{Fit to the ISR-Tevatron Experimental Data}
\par
We have studied three models, with different parameterization of $\Omega_{i,k}$, 
which were 
compared with the global experimental data base. The models are based on
the general formulae given in \eq{2CHM1} - \eq{radius}, with different
input assumptions.
Note that the fit has, in addition to the
contribution in the form of \eq{omega}, also a secondary Regge sector
(see Ref.\cite{1CH,2CH}).
This is necessary as the data base contains many experimental points
from lower ISR energies.
A study of the vacuum component alone, without a Regge contribution, is not
possible at this time, since the corresponding high energy sector of the
available data base is too small to constrain the fitted  parameters of
\eq{omega}. Unlike the assumed Pomeron exchange in the vacuum channel,
the existence of the secondary Regge trajectories is well established
both theoretically and experimentally.
As the goal of this paper is to obtain predictions in the
LHC and Cosmic Rays energy range,
where the contribution of the secondary Regge trajectories is negligibly small,
the Regge parameters are not quoted in this paper, and will be given in a 
separate publication. Note that at the Tevatron the Regge sector 
contribution is less than 1$\%$.  

\subsection{Model A}
\par
Model A is a two amplitude model, which was considered in
Ref.\cite{2CH} in detail. Its main assumption is that the
double diffraction cross section
is small enough to be omitted from the fitted data base.
As we saw, this allows us to express $\Omega_{2,2}$
in terms of $\Omega_{1,1}$ and $\Omega_{1,2}$, 
as such, this model breaks Regge factorization.
We obtain (see Refs.\cite{2CH,heralhc})
\bea
a_{el}(s,b )=i \Lb 1 - \exp\Lb - \frac{\Omega_{1,1}(s,b)}{2}\Rb-
2\beta^2 \exp\Lb - \frac{\Omega_{1,1}(s,b)}{2}\Rb
\Lb 1 - \exp\Lb - \frac{\Delta \Omega(s,b)}{2}\Rb \Rb \Rb,  \label{A1}
\eea
\bea
a_{sd}(s,b)= - i\alpha\beta \exp\Lb - \frac{\Delta \Omega(s,b)}{2}\Rb
\Lb 1 - \exp\Lb - \frac{\Delta \Omega(s,b)}{2}\Rb\Rb ,
\label{A2}
\eea
where, $\Delta \Omega = \Omega_{1,2} - \Omega_{1,1}$.
Following Ref.\cite{2CH},
we assume both $\Omega_{1,1}$ and $\Delta \Omega$ to be Gaussians in $b$.
\bea
\Omega_{1,1}(s,b) = \frac{\sigma^0_{1,1}}{\pi R^2_{1,1}(s)}\,\Lb
\frac{s}{s_0}\Rb^{\Delta} \exp\Lb - \frac{b^2}{R^2_{1,1}(s)}\Rb ,\label{A4}
\eea
\bea
\Delta \Omega(s,b) = \frac{\sigma^0_{\Delta}}{\pi R^2_{\Delta}(s)}\Lb
\frac{s}{s_0}\Rb^{\Delta} \exp\Lb - \frac{b^2}
{R^2_{\Delta}(s)}\Rb .\label{A5}
\eea
Note that in this two amplitude model 
$R^2_{\Delta} = \frac{1}{2}R_{0;11}^2  + 4\alpha^{\prime}ln (s/s_0)$
is the radius of $\Delta \Omega(s,b)$. 
We have also studied a two amplitude model in which both $\Omega_{1,1}$
and $\Omega_{1,2}$ are Gaussians in $b$. The output obtained in this two
amplitude model is compatible with Model A.
This is not surprising as $\sigma^0_{\Delta} \,\gg\,\sigma^0_{1,1}$ 
(see Table 1). 

\subsection{Model B}
\par
In the three amplitude model we do not make any assumptions
regarding the values of the double diffraction cross sections\cite{DDD} 
which are contained in our fitted data base.
We use the formulae of \eq{omega} and \eq{radius} to parameterize the three
opacities $\Omega_{1,1}$, $\Omega_{1,2}$ and $\Omega_{2,2}$,
which are taken to be Gaussians in $b$. If we assume the soft Pomeron to
be a simple fixed J pole  
the knowledge of two out of those three opacities   determines
 the third. We denote this option Model B(1). In this model
$\sigma^0_{1,2} = \sqrt{\sigma^0_{1,1}\times \sigma^0_{2,2}}$. As we shall see,
the fit corresponding to Model B(1) is not satisfactory.
We have, also, studied Model B(2) in which a  factorization of
$\sigma^0_{i,k}$ is not assumed. Accordingly, $\sigma^0_{1,1}$, 
$\sigma^0_{1,2}$
and $\sigma^0_{2,2}$ are independent fitted parameters of the model. 
This model reflects our approach to the soft interaction as the 
continuation of the hard processes 
in the saturated soft region. We would like to stress, that
this model gives a very good reproduction of the data.
\subsection{Results}
\par
The parameters of Model A are based on
a fit to a 55 experimental data points base
which includes the $p$-$p$ and $\bar{p}$-$p$ total cross sections,
integrated elastic cross sections,
integrated single diffraction cross sections,
and the forward slope of the elastic cross section
in the ISR-Tevatron energy range.
We did not include the Cosmic Ray air showers estimated
total cross sections in our data base, as
they  require additional  model dependent assumptions\cite{Block}.
As stated, we neglected the reported DD cross section points.
The fitted parameters of Model A are listed in Table 1
with a corresponding $\chi^2/(d.o.f)$ of 1.50.
\begin{footnotesize}
\TABLE[ht]{
\begin{tabular}{|l|l|l|l|l|l|l|l|l|c|}
\hline
Model & $ \Delta $ & $\beta$ &  $R^{2}_{0;1,1}$  & $\alpha^{\prime}_{P}$&
$\sigma^0_{1,1}$ &
$\sigma^0_{2,2}$ & $\sigma^0_{\Delta}$ & $\sigma^0_{1,2}$ \\ \hline
A & 0.126 & 0.464 &16.34 $GeV^{-2}$ &0.200 $GeV^{-2}$ &
12.99 $GeV^{-2}$& N/A & 145.6 $GeV^{-2}$& N/A \\
\hline
B(1)& 0.150 & 0.526 &20.80 $GeV^{-2}$& 0.184 $GeV^{-2}$ &4.84 $GeV^{-2}$&
4006.9 $GeV^{-2}$& N/A  &139.3 $GeV^{-2}$\\
\hline
B(2) &   0.150 & 0.776 &20.83 $GeV^{-2}$& 0.173 $GeV^{-2}$ &9.22 $GeV^{-2}$&
3503.5 $GeV^{-2}$& N/A &6.5 $GeV^{-2}$ \\
\hline
\end{tabular}
\caption{Fitted parameters for Models A, B(1) and B(2).}}
\end{footnotesize}
\par
The fits to Models B(1) and B(2) are based on an updated data base which
includes the data base used for Model A, plus 5
double diffraction cross sections data points\cite{DDD}.
In Table 1 we present two sets of fitted parameters for Model B.
The factorizable Model B(1) does not give a good reproduction of the
data,  with a $\chi^2/(d.o.f.)$=2.30. Part, but not all, of this large
$\chi^2$
is contributed by the DD data, where the model predictions are
significantly
below the experimental points.
Model B(2), with a $\chi^2/(d.o.f.)$ = 1.25, provides a
very good reproduction of the data base.
Its seemingly high $\chi^2$ reflects the poor quality of the
published SD data points.
The model provides a very good reproduction of the DD data points.
\DOUBLEFIGURE[ht]{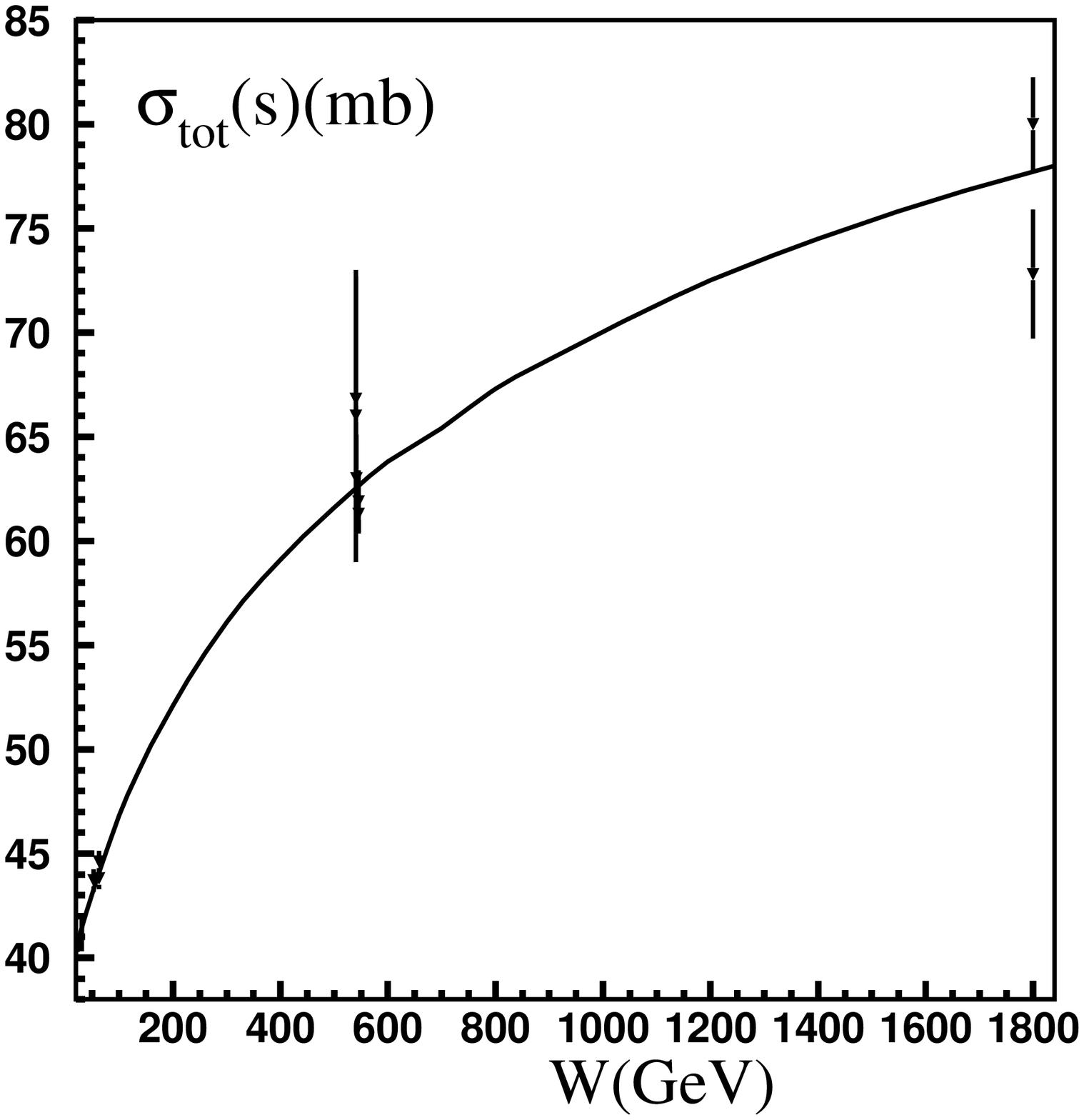,width=90mm,height=80mm}{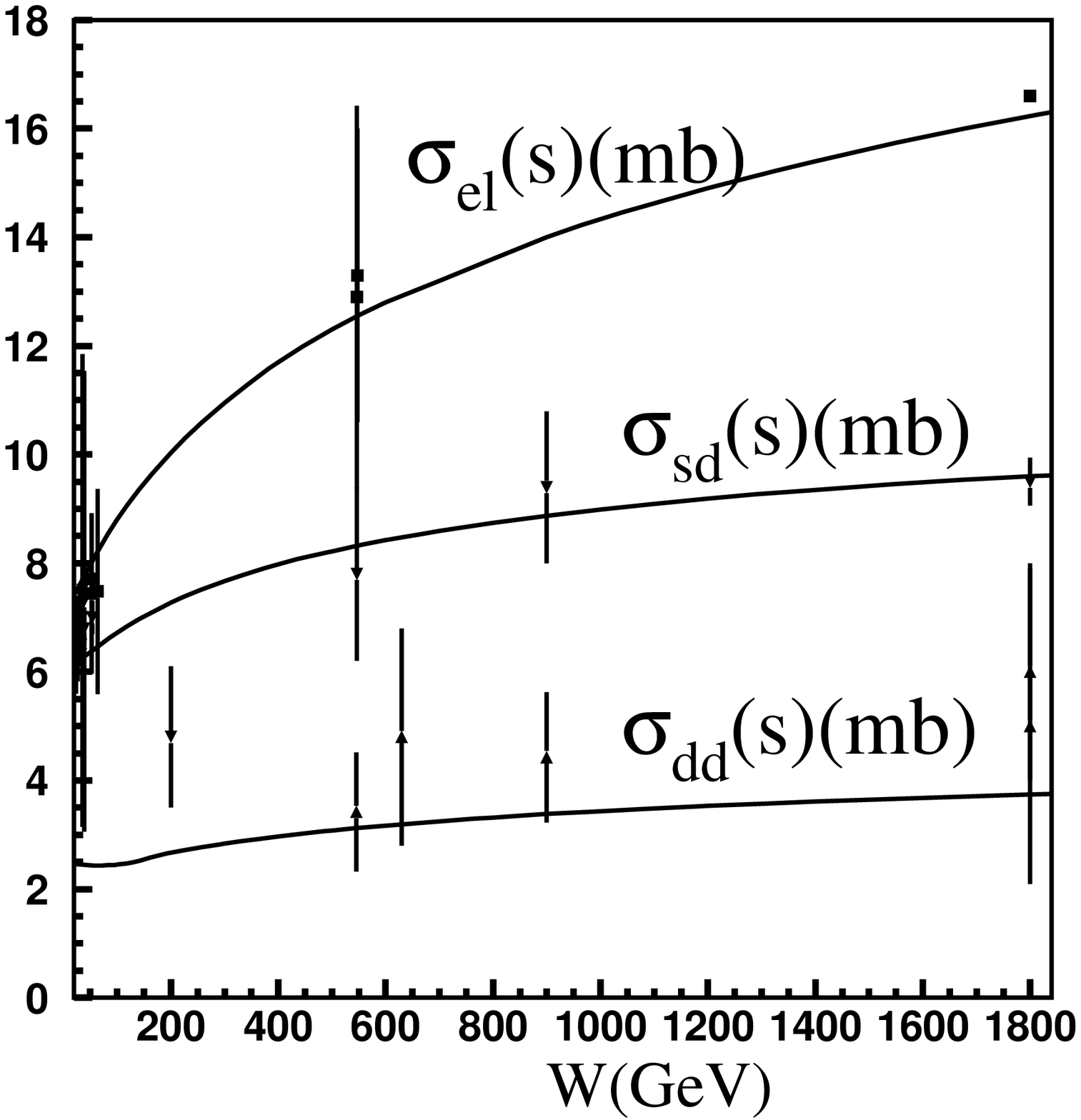,width=90mm,height=80mm}
{UA4-Tevatron energy dependence of $\sigma_{tot}$ in Model B(2).
\label{xstot}}
{UA4-Tevatron energy dependence of $\sigma_{el}$, $\sigma_{sd}$ and $\sigma_{dd}$
in Model B(2).
\label{xsel}}
\par
Model A and Model B(2)
give compatible reproductions of our data base without the DD points.
To this end, we refer to the figures of Ref.\cite{2CH} and
\fig{xstot} and \fig{xsel} of this paper.
The calculations of $B_{sd}$ and $\rho$
were executed with the fitted parameters of Model B(2). Neglecting the Regge
contribution, we reproduce the higher energy
experimental data points obtained by UA4, CDF and E710.
\par
Obviously, factorization violations in the coupling input of Model A and
Model B(2) are very interesting and support our partonic saturation approach.
We hope that our present
results will encourage additional studies of the intriguing relations
between, what are traditionally called, soft and hard Pomerons.

\section{Predictions for LHC and Cosmic Ray Energies}
Model B(2) cross section and slopes 
predictions for LHC and Cosmic Ray energies are 
summarized in \fig{spsd}, \fig{slopes} and Table 2.
At W=14 $TeV$ (LHC energy) our predicted cross sections are: 
$\sigma_{tot}=110.5\,mb$, $\sigma_{el}=25.3\,mb$, $\sigma_{sd}=11.6 \,mb$ 
and $\sigma_{dd}=4.9\,mb$.
These predictions are slightly
higher than those obtained\cite{2CH} in Model A. 
The corresponding forward slopes are: $B_{el}=20.5 \,GeV^{-2}$, 
$B_{sd}=15.9\,GeV^{-2}$ and $B_{dd}=13.5\,GeV^{-2}$.
One can see that the predicted cross section of the diffractive channels, 
as compared with the elastic cross section, is relatively large
and should be considered in the estimates for the
background for diffractive Higgs production process.
The ratio of 
$\sigma_{el}/\sigma_{tot}$ at the LHC is predicted to be less than 0.25. 
This is a signature that at the LHC energy the proton-proton (anti proton) 
elastic scattering amplitude is well below
the black disc asymptotic bound. We shall elaborate on this point in 
Sec. 6.
\FIGURE[ht]{
\centerline{\epsfig{file= 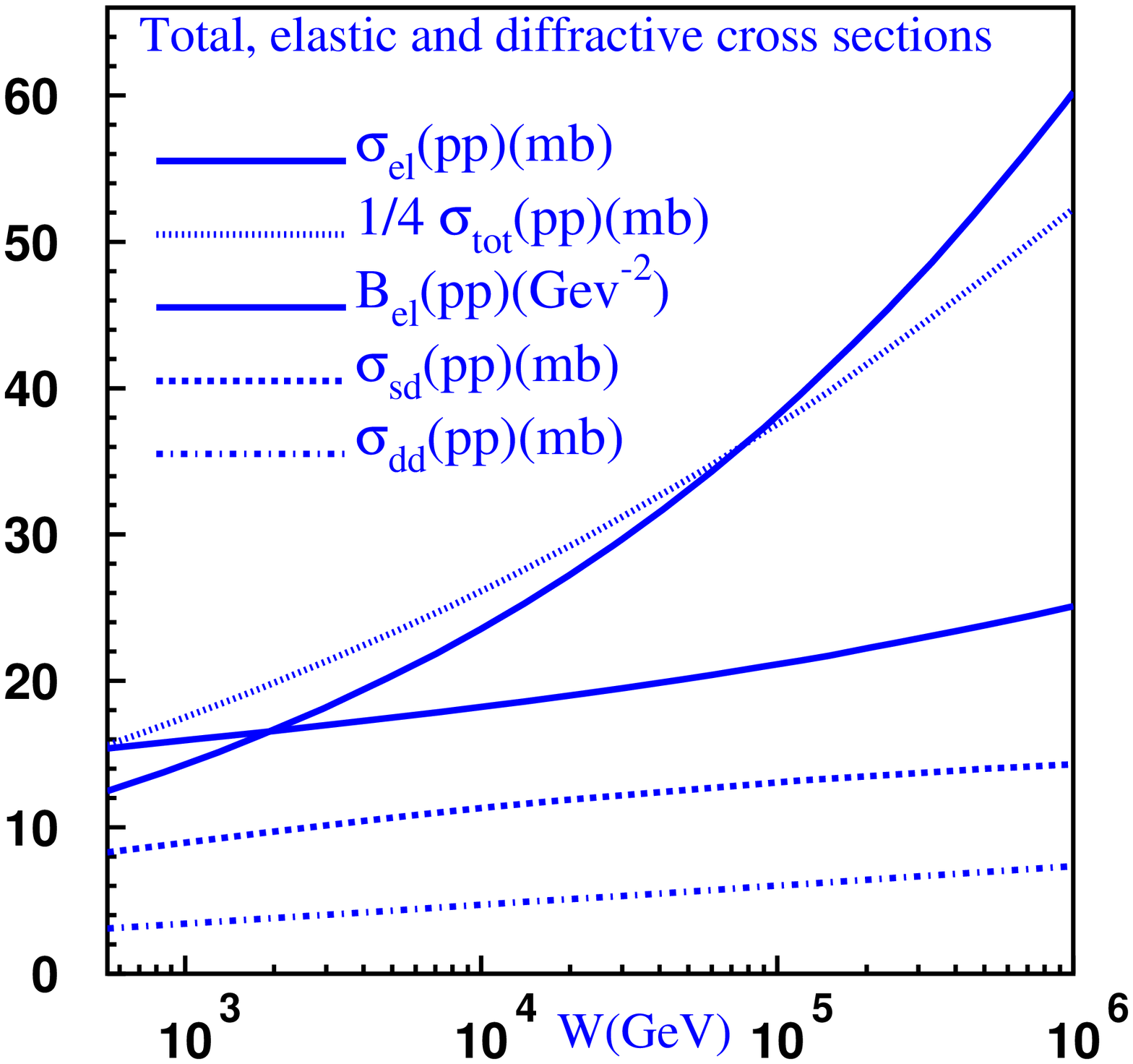,width=140mm,height=120mm}}
\caption{Total, elastic and diffractive cross sections in Model B(2).}
\label{spsd}}
\FIGURE[h]{
\centerline{\epsfig{file=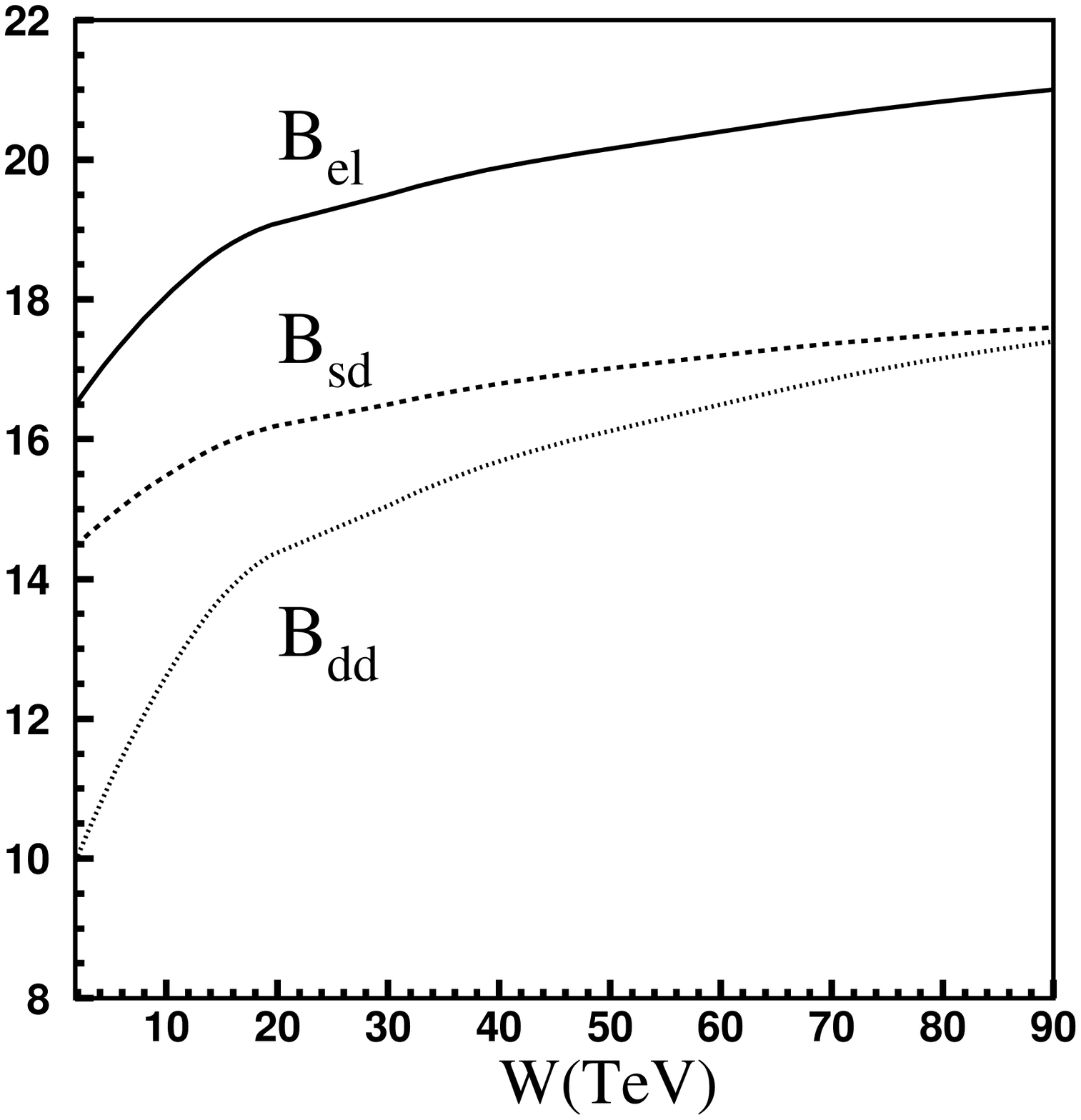,width=120mm,height=100mm}}
\caption{Energy dependence of $B_{el}$, $B_{sd}$ and $B_{dd}$
in Model B(2).}
\label{slopes}}

\begin{footnotesize}
\TABLE[ht]{
\begin{tabular}{|l|l|l|l|l|l|l|c|c|c|c|}
\hline 
&         &                       &                
&         &                       &                
&         &                       &\\                        
$\sqrt{s}$& $ \sigma^{DL}_{tot} $ & $ \sigma_{tot} $
&  $  \sigma_{el} $  & $ \sigma_{sd} $&
$\sigma_{dd}$& $B_{el}$ & $R_{el}$ & $R_D$
& $\frac {\sigma_{diff}} {\sigma_{el}}$ \\
TeV & mb & mb & mb & mb & mb & $GeV^{-2}$&  &  & \\ \hline
1.8 & 73.0 & 78.0 & 16.3 & 9.6 & 3.8 & 16.8 & 0.21 & 0.38 &0.83\\
14 & 101.7 & 110.5 & 25.3 & 11.6 & 4.9 & 20.5 & 0.23 & 0.38 &0.65 \\
30 &115.0 & 124.8 & 29.7 & 12.2 & 5.3 & 22.0 & 0.24 & 0.38 &  0.59\\
60 & 128.6 & 139.0 & 34.3 & 12.7 & 5.7 & 23.4 & 0.25 & 0.38 &0.54 \\
120 & 143.9 & 154.0 & 39.6 & 13.2 & 6.1 & 24.9 & 0.26 & 0.38 & 0.49\\
250 & 162.0 & 172.0 & 45.9 & 13.6 & 6.6 & 26.5 & 0.27 & 0.38 & 0.44 \\
500 & 181.2 & 190.0 & 52.7 & 14.0 & 7.0 & 28.1 & 0.28 & 0.39 & 0.40 \\
1000 & 202.7 & 209.0 & 60.2 & 14.3 & 7.4 & 29.8 & 0.29 & 0.39 & 0.10 \\
$10^{11}$ & 3970.0 & 1070.0 & 451.2 & 21.6 & 19.5& 109.9 & 0.42 & 0.46 & 0.09\\
1.22\,$10^{19}$ & 26400.0& 1970.0 & 871.4 & 25.5 &27.7 & 202.6 & 0.44 & 0.47 & 
0.06\\
(Planck) & & & & & & & & & \\
\hline 
\end{tabular}
\caption{Cross sections and elastic slope in Model B(2). $\sigma^{DL}_{tot}$ is 
presented for comparison.}}
\end{footnotesize}
\par
Checking the predictions presented in Table 2, we note a systematic
behavior of the listed cross sections from which we learn about 
the gross features of unitarity, spanning the energies from accelerator 
range to the Planck mass.  
\newline
1) The difference between DL non screened total cross section 
predictions\cite{DL} and GLM is small up to the GZK ankle, which 
is a practical upper bound for Cosmic Ray energies with which we can obtain 
information relevant to our study. 
\newline
2) The predicted ratio of $R_{el}$ and $R_D$ are well below 0.5 up to
GZK and above. Close to the Planck mass the ratios approach 0.5.  
\newline
3) From Table 2 we see that while $R_{el}$ grows very slowly with
energy, $R_D$ 
is essentially a constant, slightly less than 0.4, all through the Cosmic 
Ray energy range.
Accordingly, the predictions for the diffractive channels, though 
increasingly suppressed with energy relative to
the elastic channel, cannot be ignored. 
\newline
4) Well above the GZK cutoff the ratio $R_{el}$ grows more rapidly with energy 
while $\sigma_{diff}/\sigma_{el}$ which is less than 0.1, diminishes slowly. 
At the Planck mass we see that $R_{el}$=0.44 and $R_D$=0.47.
\newline
5) The above predictions suggest a very slow
onset of unitarity constraints on the total and integrated cross sections
with growing energy. We wish to remind the reader
that in eikonal models the asymptotic behavior 
of $\sigma_{tot}$, with a logarithmic accuracy, 
is $ln^2(s/s_0)$, that of $\sigma_{el}$ is $\frac{1}{2} ln^2(s/s_0)$,
while that of $\sigma_{diff}$ is only $ln(s/s_0)$.
This issue will be discussed in detail in Sec. 6.

\section{Survival probabilities in the GLM model}
\par
In the following we shall limit our discussion to the survival probability
of Higgs production in an exclusive
central diffractive process, calculated in a three amplitude model.
A general review of
survival probability calculations can be found in
Ref.\cite{heralhc}. Our one and two amplitude model
calculations have been published\cite{SP1CH,SP2CH}.
\par
In our model we assume an input Gaussian $b$-dependence
for both the elastic opacities, specified in \eq{omega} - \eq{radius},
and the similar structured hard diffractive amplitude of interest.
The hard diffractive non screened cross sections
in the (i,k) channel are calculated
using the multi particle optical theorem\cite{Mueller}.
As stated, they are written in the same form as the soft amplitudes
\begin{equation}\label{3.6}
{\Omega_{i,k}^H}={\nu_{i,k}^H(s)} \Gamma_{i,k}^H(b),
\end{equation}
where, 
\begin{equation}\label{3.7}
\nu_{i,k}^H=\sigma_{i,k}^{H0}(\frac{s}{s})^{\Delta_H},  
\end{equation}
\begin{equation}\label{3.8}
\Gamma_{i,k}^H(b)=\frac{1}{\pi R_{i,k}^2}\,e^{-\frac{\,b^2}{R_{i,k}^H}^2}.
\end{equation}
The hard radii ${R_{i,k}^H}^2$
are constants derived from HERA $J/\Psi$ elastic and inelastic
photo and DIS production\cite{KOTE,PSISL}.
See, also, Ref.\cite{heralhc}.
\FIGURE[ht]{
\centerline{\epsfig{file=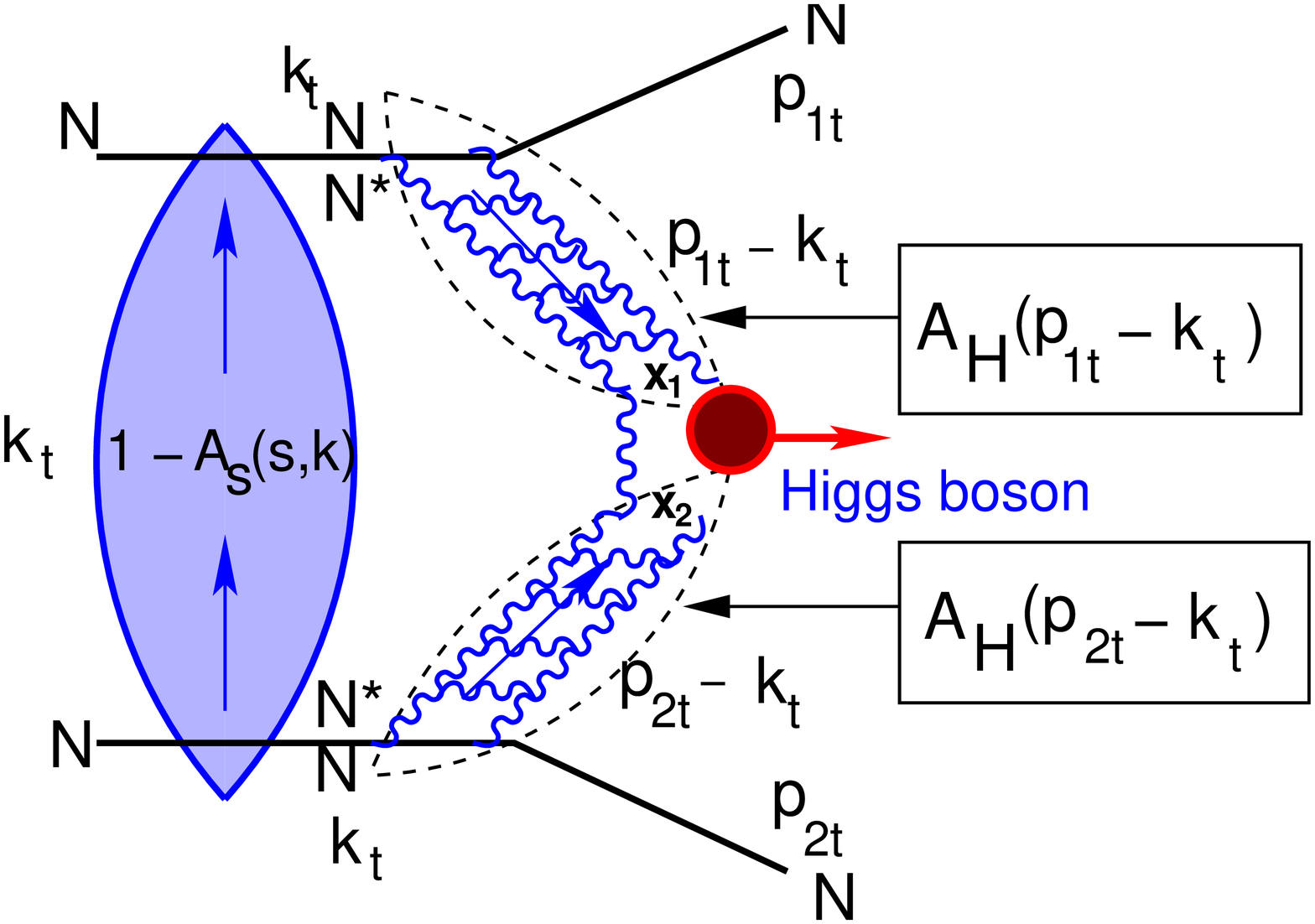,width=80mm}}
\caption{Survival probability for exclusive central diffractive 
production of the Higgs boson.} 
\label{sp-dia}}
\par
The general formulae for the calculation of the survival probability for
diffractive Higgs boson production have been discussed
in Refs.\cite{SP2CH,heralhc,SP3P}. The structure of the survival probability
expression is shown in \fig{sp-dia}. Accordingly,
\beq \label{SP}
\SP = \frac{N(s)}{D(s)},
\eeq
where,
\bea
&N(s) = \int d^2\,b_1\,d^2\,b_2
\left[A_H(s,b_1)\,A_H(s,b_2) (1-A_S \Lb(s,(\mathbf{b}_1+\mathbf{b}_2 )
\Rb)\right]^2, \label{SP1}\\
&D(s) = \int\,d^2\,b_1\,d^2\,b_2
\left[A_H(s,b_1)\,A_H(s,b_2)\right]^2.
\label{SP2}
\eea
$A_s$ denotes the soft strong interaction amplitude given
by \eq{2CHM1}.
The form of $A_H(s,b)$ has been discussed in detail in Refs.\cite{SP2CH,heralhc}.
\par
Using \eq{EL}-\eq{DD}, the integrands of
\eq{SP1} and \eq{SP2} are reduced by eliminating common $s$-dependent
expressions.
\bea
&&N(s) = \int\,d^2 b_1 d^2 b_2 
[A_H(s,b_1)\,A_H(s,b_2) (1 - A_S \Lb
\mathbf{b} = \mathbf{b}_1+\mathbf{b}_2 \Rb)]^2 \nonumber \\
&& = \int d^2 b_1 d^2 b_2\,
[(1 - a_{el}(s,b))A^{pp}_H(b_1) A^{pp}_H(b_2)
 - a_{sd}(s,b)\Lb A^{pd}_H(b_1) A^{pp}_H(b_2) + 
A^{pp}_H(b_1) A^{pd}_H(b_2) \Rb \nonumber \\
&& - a_{dd}(s,b) A^{pd}_H(b_1) A^{pd}_H(b_2)]^2
\label{SP3},
\eea
\beq \label{SP4}
D = \int d^2 b_1 d^2 b_2 
\left[A^{pp}_H(b_1) A^{pp}_H(b_2)\right]^2.
\eeq
\par
Following Refs.\cite{SP3P,heralhc} we introduce 
two hard $b$-profiles
\bea
A^{pp}_H(b) &=&
\frac{V_{p \to p}}{2 \pi B_{el}^H} \exp \Lb-\frac{b^2}{2\,B_{el}^H} \Rb,
\label{2C10}\\
A^{pd}_H(b) &=& \frac{V_{p \to d}}{2 \pi
B_{in}^H} \exp \Lb -\frac{b^2}{2 B_{in}^H}\Rb.
\label{2C11}
\eea
The values $B_{el}^H$=3.6 $GeV^{-2}$ and $B_{in}^H$=1 $GeV^{-2}$
have been taken from the experimental HERA data on
$J/\Psi$ production in HERA (see Refs.\cite{KOTE,PSISL}).
\par
\FIGURE[h]{
\centerline{\epsfig{file= 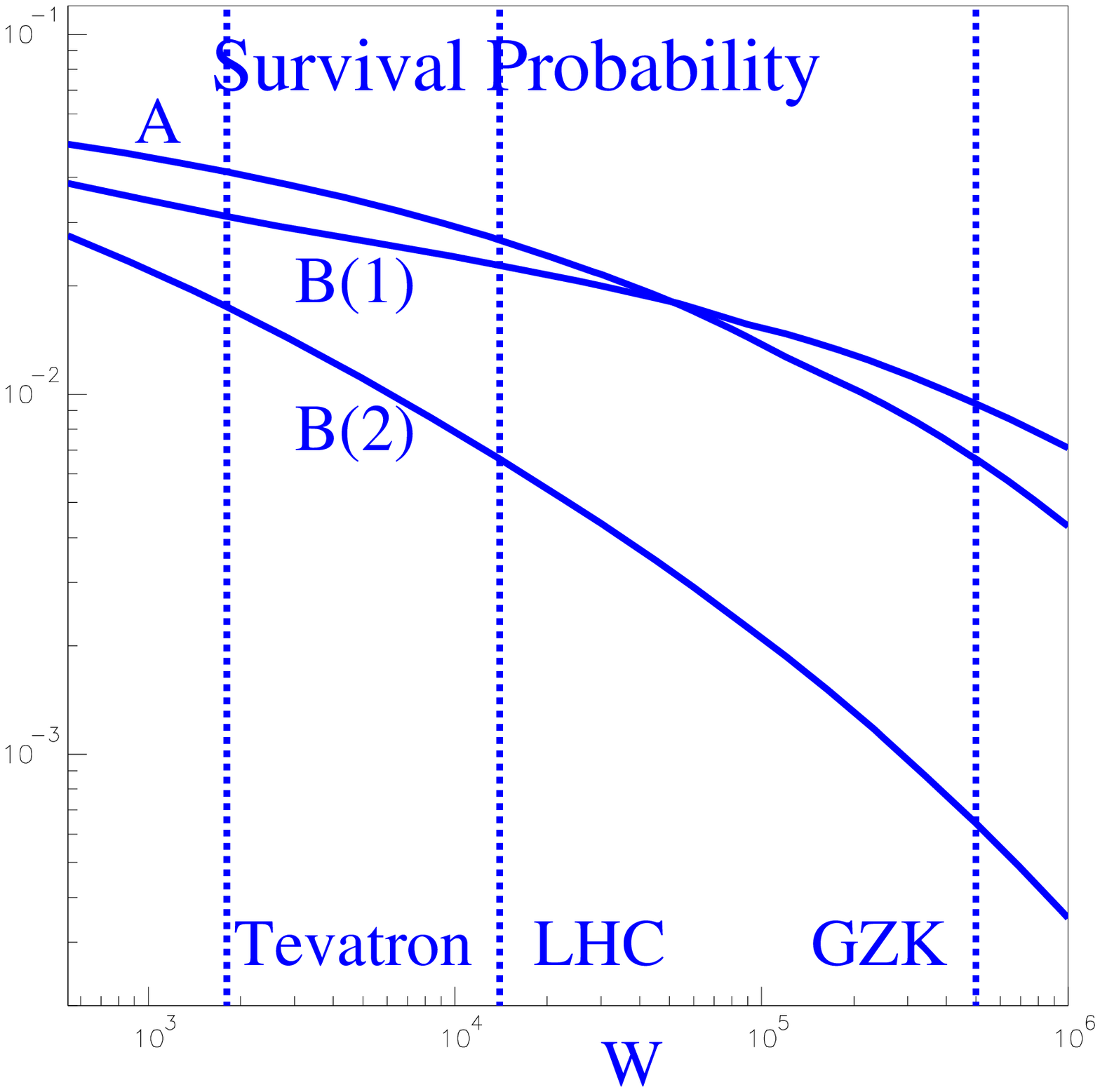,width=120mm,height=100mm}}
\caption{Energy dependence of centrally produced Higgs survival probability
calculated in Models A, B(1), B(2).}
\label{sp}}
\par
The calculated survival probabilities have been instrumental in the theoretical
interpretation of hard LRG di-jets produced at the Tevatron. For details see
Ref.\cite{heralhc}.
Using \eq{SP}-\eq{SP4} we calculate the survival probability $S^2_H$ for
exclusive Higgs production in central diffraction. Our results
are plotted in \fig{sp}.
\par
In the following we focus on our LHC predictions based on the above. 
$S^2_H$ for exclusive Higgs production in central diffraction has been
calculated\cite{heralhc} in the two amplitude Model A. The resulting
$S^2_H\,=\,0.027$ is essentially the same as
the predictions of the Durham group\cite{KKMR} 
and Frankfurt et al.\cite{S2FS}. 
The main conclusion of our present results,
obtained in the three amplitude B(1) and B(2) Models, is that opening more 
screening rescattering channels results in a reduction of the calculated value of
$S^2_H$. Its LHC value in Model B(1) is 0.02 and in
Model B(2) it is 0.007.
This last result has also been obtained in Ref.\cite{S2FS}.
Some clarifications concerning the reliability of our $S^2_{H}$
estimates are in order. Our model provides am excellent reproduction of the
elastic amplitude for $b > 1.5 - 2.0 fm$. However, since we do not 
reproduce 
$d \sigma_{el}/d t$ outside the forward cone  well \cite{Rlet}, it would
appear that  
our reproduction of $a_{el}$ at
small $b$ is deficient.  
However, we note that a very good reproduction of $d\sigma_{el}/dt$ at 
$|t|\,>\,0.2\,GeV^2$ is obtained in eikonal
models using dipoles or multi-poles in $t$-space for the profile function.
The difference between these opacities and ours is 
very small at small b values\cite{Zur,S2FS}.  
Accordingly, the error introduced into our calculation of $S^2$ 
is estimated to be small. An actual comparison\cite{GKLMP} supports 
this conclusion.
\par
A realistic estimate of the survival probability
is crucial for the experimental discovery of a diffractive Higgs at the LHC.
An educated guess of its value
can be obtained, at an early stage of LHC operation, through a
measurement of the rate of central hard LRG di-jets production (a GJJG 
configuration) coupled to a study of its expected rate in a non screened pQCD 
calculation.

\section{The Role of Unitarity at High Energies}

\subsection{Unitarity experimental signatures}
\par
As it stands, the experimental support indicating the important role of 
unitarity considerations is confined to the energy range attained by 
existing accelerators.
 Experimental signatures associated with unitarity are observed in 
inelastic 
diffractive channels (soft and hard) at relatively low energies. 
This is a consequence of the unitarity suppression which, at a given impact 
parameter $b$, is given by $P_{i,k}(s,b)=e^{-\Omega_{i,k}(s,b)}$. 
After its convolution with the inelastic diffractive amplitude of interest  
we obtain the survival probability. 
In this context we  identify a few significant features: 
\newline
1) The experimental data shows a severe moderation of the energy dependence of 
soft diffraction at ISR energies and above, and it increase
with energy is much slower than elastic scattering. This observation 
differs from the conventional Regge expectation in which the cross 
sections of elastic and inelastic 
diffraction are expected to have a similar energy dependence.
From Table 2 we see that the ratio
$\sigma_{diff}/\sigma_{el}$ decreases monotonically with $s$ from 
low energies to the Planck mass, where it is negligible. 
In the ISR-Tevatron range this has been observed  
experimentally\cite{Dino}, and is well reproduced by Model B(2).
As we have noted our calculations at exceedingly high energy are fully 
compatible with the asymptotic behavior enforced by unitarity.
\newline 
2) The predicted small value  of $S^2_{H}$, the survival probability of 
a given LRG hard inelastic diffractive channel, has been verified in the Tevatron 
studies of LRG hard di-jets in various kinematical configurations. 
The survival probabilities are 
essential in order to understand both the relative smallness and energy 
dependence of the measured LRG hard di-jets experimental rates, when 
compared 
with pQCD estimates. Survival probabilities are essential in 
explaining\cite{KKMR} the large factorization breaking in the rates of 
LRG hard di-jets observed in Fermilab and HERA.  
The above has been systematically obtained by a few independent models, 
regardless of the  method chosen to enforce unitarity. 
For details see Ref.\cite{heralhc}.   
\newline
3) The assessment of $S_H^2$ is a fundamental ingredient in the
experimental program to discover a Higgs boson produced in central diffraction. 
Its signature is that the diffractive system,
observed in the $\eta - \phi$ final state lego plot,
is accompanied on both sides by a LRG.
In this paper we have presented Model B(2) predicted value of $S_H^2$ = 0.007, 
smaller than our Model A prediction of $S_H^2$ = 0.027. Regardless 
of our particular formulation and estimate, unitarity implies that 
the experimental rate for a diffractive Higgs, if and when it will be 
discovered, is bound to be approximately two orders of magnitude smaller 
(order of 1\%) than the pQCD estimate.
\subsection{An amplitude analysis of unitarity at very high energies}
\begin{footnotesize}
\TABLE[ht]{
\begin{tabular}{|c|c|c|}
\hline
$W=\sqrt{s}$ & $a_{el}(b = 0)$ & $a_{el}$ black \\
           &                  & core radius \\
   $TeV$   &                  & fm \\
\hline
1.8  &0.62 & \\
14 & 0.71 & \\
60 & 0.79  & \\
250 & 0.87 & \\
500 & 0.90 & \\
$10^3$ & 0.93 &  \\
3 $10^4$& 1.00 & 0.5  \\
6 $10^4$& 1.00 & 0.8  \\
3 $10^8$& 1.00 & 2.3  \\
$10^{11}$ & 1.00 & 3.0  \\
1.22 $10^{19}$ & 1.00 & 4.6  \\
(Planck)& & \\
\hline
\end{tabular}
\caption{Impact parameter behaviour of $a_{el}(s,b = 0)$. }}
\end{footnotesize}
\par
The basic amplitudes of the GLM model 
are $A_{1,1}$, $A_{1,2}$, and $A_{2,2}$
whose $b$ structure is specified in \eq{2CHM1}). 
These are the building blocks
with which we construct $a_{el}$, $a_{sd}$, 
$a_{dd}$, (\eq{EL}-\eq{DD}). 
The $A_{i,k}$ amplitudes are bounded by the  black
disc unitarity bound. 
Checking Table 1, it is evident that $\Omega_{2,2}$
is  considerably larger than 
$\Omega_{1,1}$ and $\Omega_{1,2}$.
As a consequence, the amplitude 
$A_{2,2}(s,b)$ reaches
the unitarity bound of 1 at very low energies. 
This blackness extends to higher 
$b$ values with increasing energy. 
The observation that $A_{2,2}(s,b)$=1 at 
given values of $(s,b)$ does not 
imply that the physical elastic scattering amplitude 
has reached the unitarity bound at these 
$(s,b)$ values, see \fig{ampab25} 
where we show both the basic amplitudes 
$A_{i,k}$, as well as, 
$a_{el}$, $a_{sd}$ and $a_{dd}$ at W=25 $GeV$.
$a_{el}(s,b)$ reaches the black disc bound when
$A_{1,1}(s,b)$=$A_{1,2}(s,b)$=$A_{2,2}(s,b)$=1.
Accordingly,
$a_{el}(s,b)$=1 and $a_{sd}(s,b)$=$a_{dd}(s,b)$=0.
This result is independent of the fitted value
of  $\beta$.
A fundamental feature of the GLM models is that
$a_{el}$ approaches the black
disc bound at small $b$ very slowly,
reaching it at approximately W=30,000 $TeV$.
See Table 3 and \fig{ampel}.

\FIGURE[ht]{\begin{tabular}{c c}
\epsfig{file=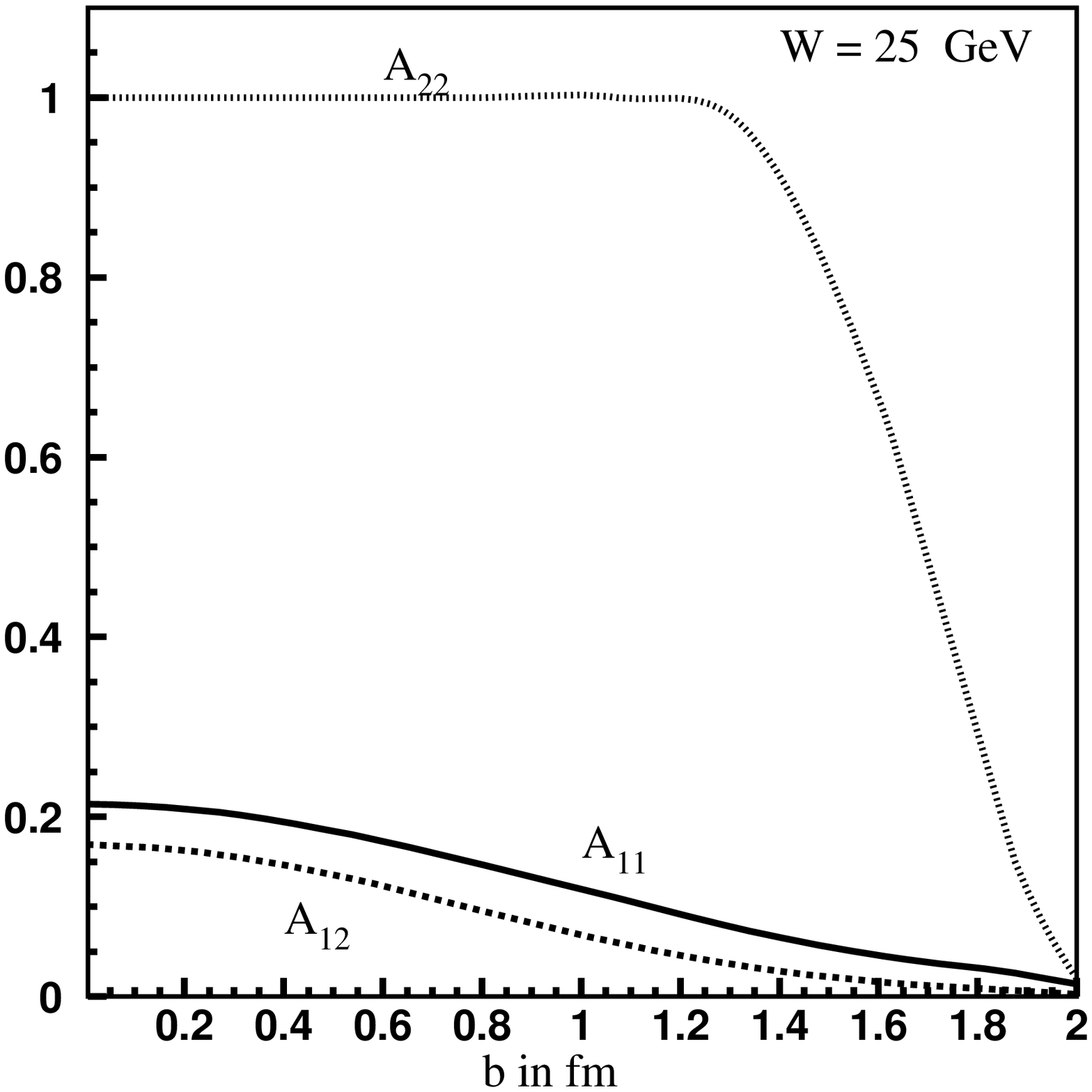,width=90mm,height=80mm}  &
\epsfig{file=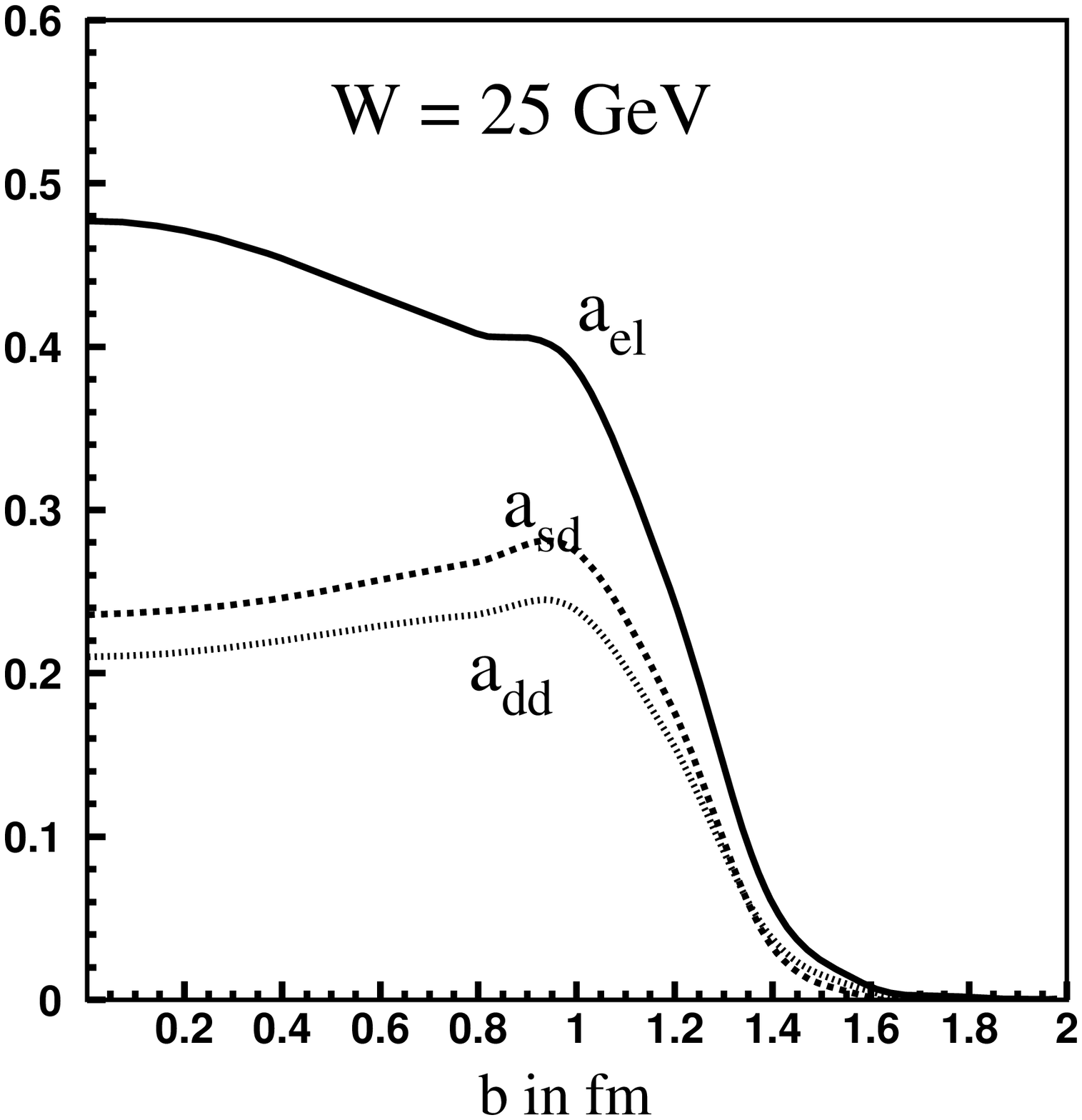,width=90mm,height=80mm} \\
\end{tabular}
\caption{Impact parameter dependence of $A_{i,k}$ and 
$a_{el}$, $a_{sd}$, $a_{dd}$ in Model B(2) at W = 25 $GeV$.
\label{ampab25}}}

\FIGURE[ht]
{\epsfig{file=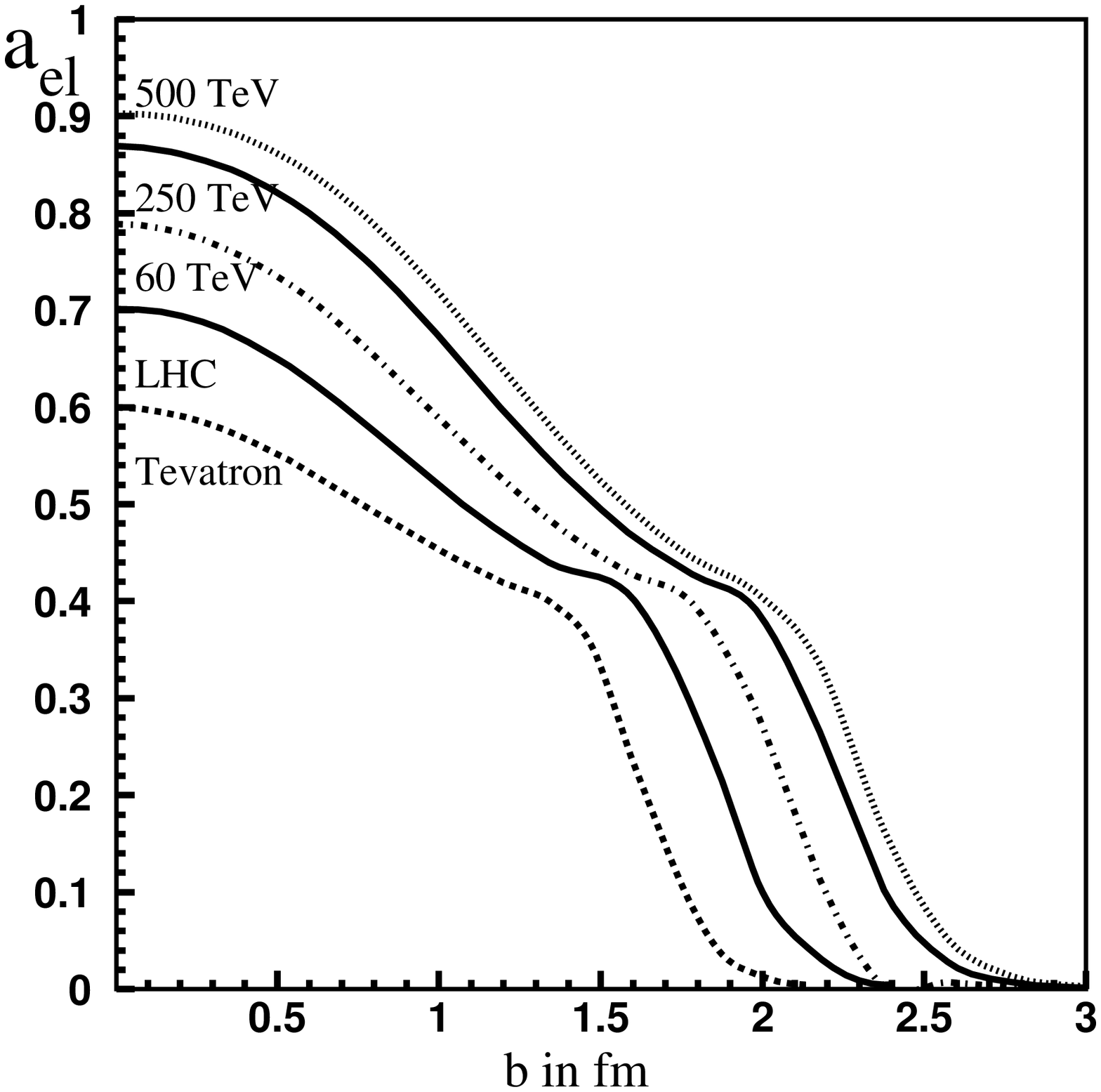,width=140mm}
\caption{Impact parameter dependence of $a_{el}$ 
in Model B(2) at different energies.}
\label{ampel}}

\FIGURE[ht]{\begin{tabular}{c c}
\epsfig{file=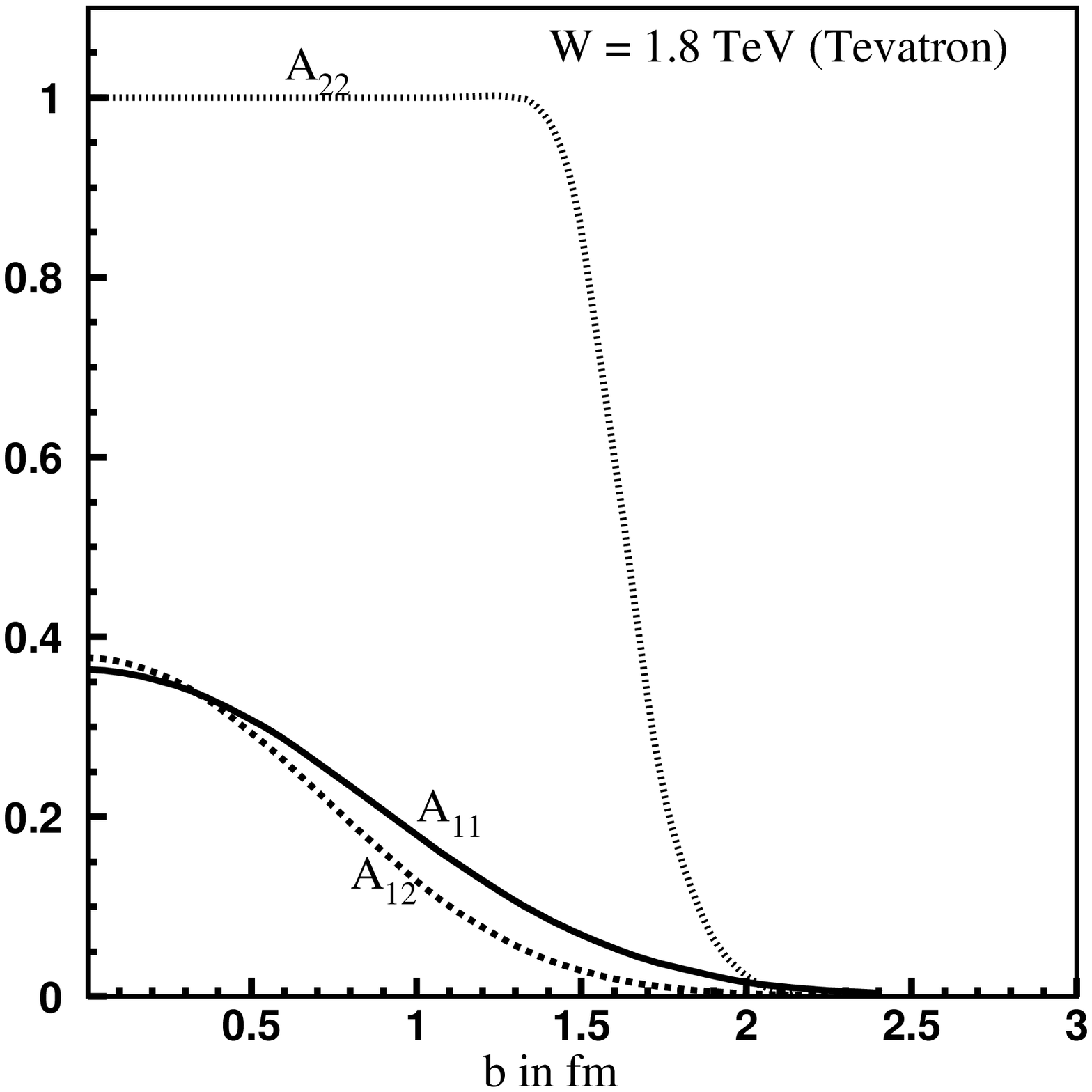,width=90mm,height=80mm} & 
\epsfig{file=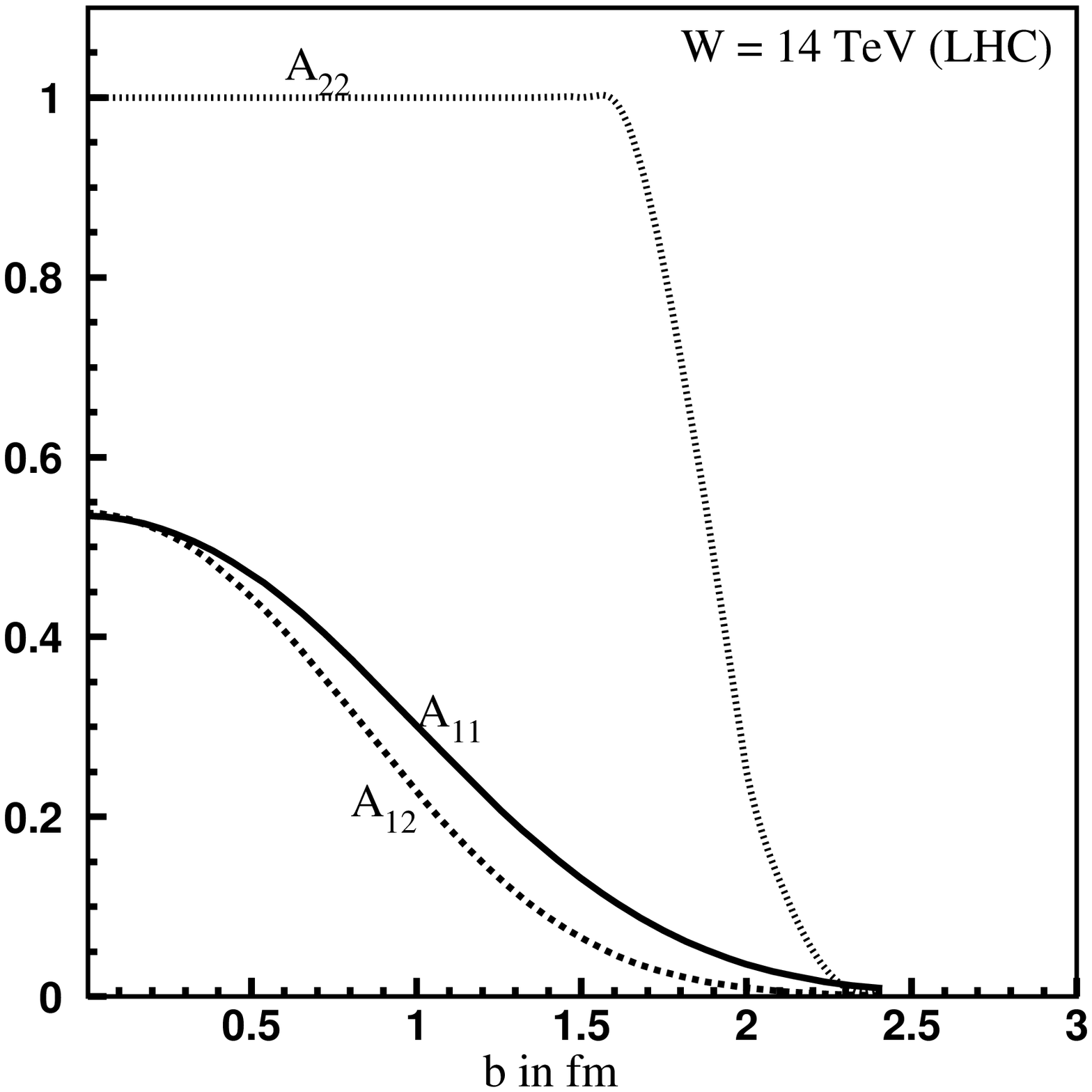,width=90mm,height=80mm} \\
\end{tabular}
\caption{Impact parameter dependence of $A_{i,k}$
at the Tevatron and LHC in Model B(2).}
\label{ampb}}

\FIGURE[ht]{\begin{tabular}{c c}
\epsfig{file=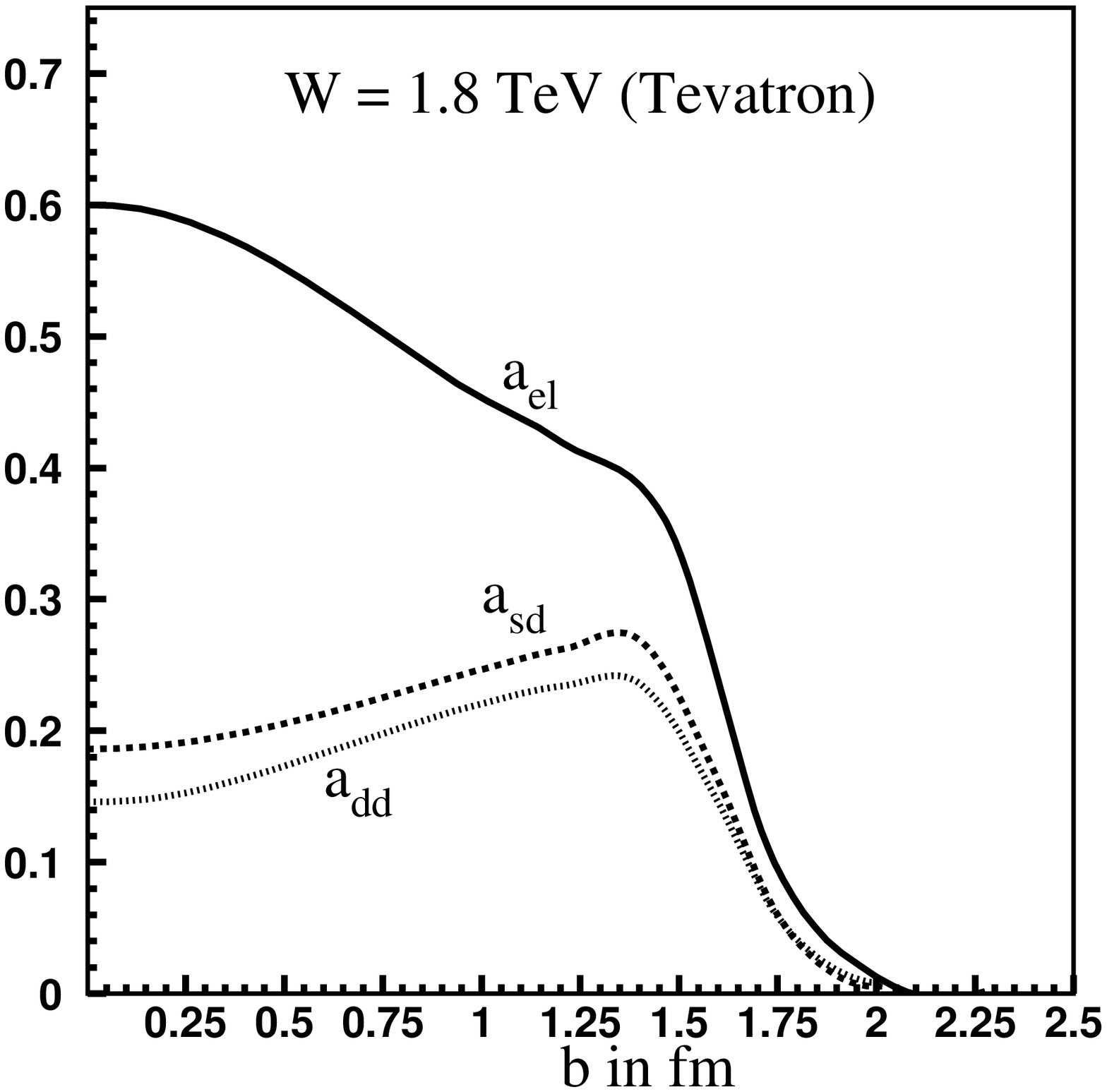,width=90mm,height=80mm} &
\epsfig{file=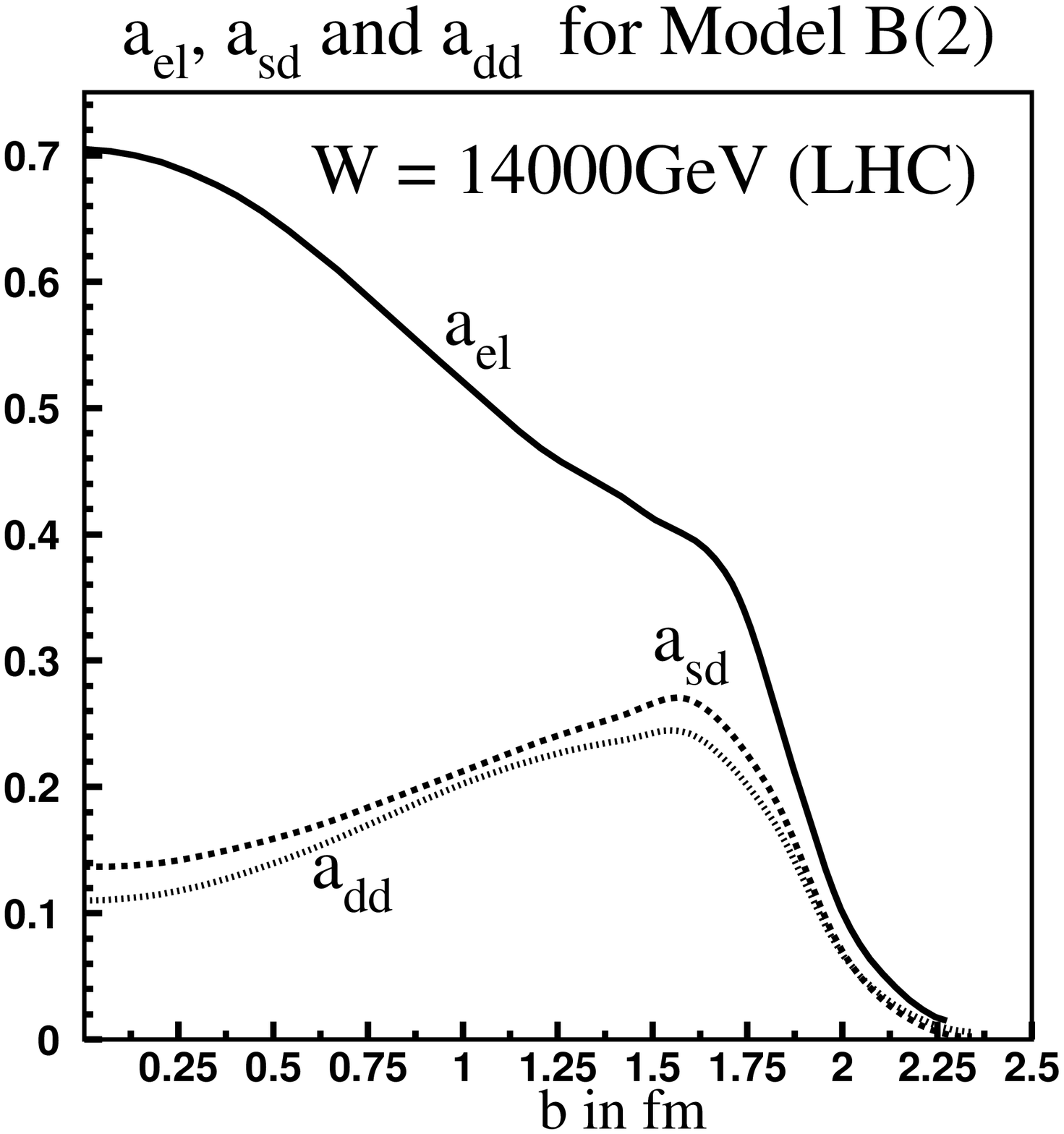,width=90mm,height=80mm} \\
\end{tabular}
\caption{Impact parameter dependence of 
$a_{el}$, $a_{sd}$ and $a_{dd}$ at the Tevatron and LHC in Model B(2).}
\label{amlallb}}
\par
A consequence of $\Omega_{i,k}$ being central in $b$, is that
$P_{i,k}^S(s,b)$ is very small at $b$ = 0 and 
monotonically approaches 
its limiting value of 1, in the high $b$ limit.
As a result, given an input (non screened) diffractive
amplitude which is central in $b$, the output (screened)
diffractive amplitude is peripheral.
This important feature, as calculated in Model B(2), is shown
in \fig{ampb} and \fig{amlallb} at the Tevatron and LHC energies.
As expected, the diffractive amplitudes have a local
minimum at $b$=0, 
and a maximum at $b \approx 1.4 fm$ for the Tevatron, which moves to
higher values of $b$ with the growth of energy.
This implies a non trivial $t$ dependence of
$d\sigma_{diff}(M^2_{diff})/dt$ in the diffractive channels.
These qualitative features are induced by Model A, Model B(1) and Model B(2)
considered in this paper, even though their detailed behavior, as seen in
\fig{amsdbm}, are not identical.
Experimentally, this feature is not easily tested 
as it requires a fine grid of $M_{diff}^2$.

\FIGURE[ht]{\begin{tabular}{c c}
\epsfig{file=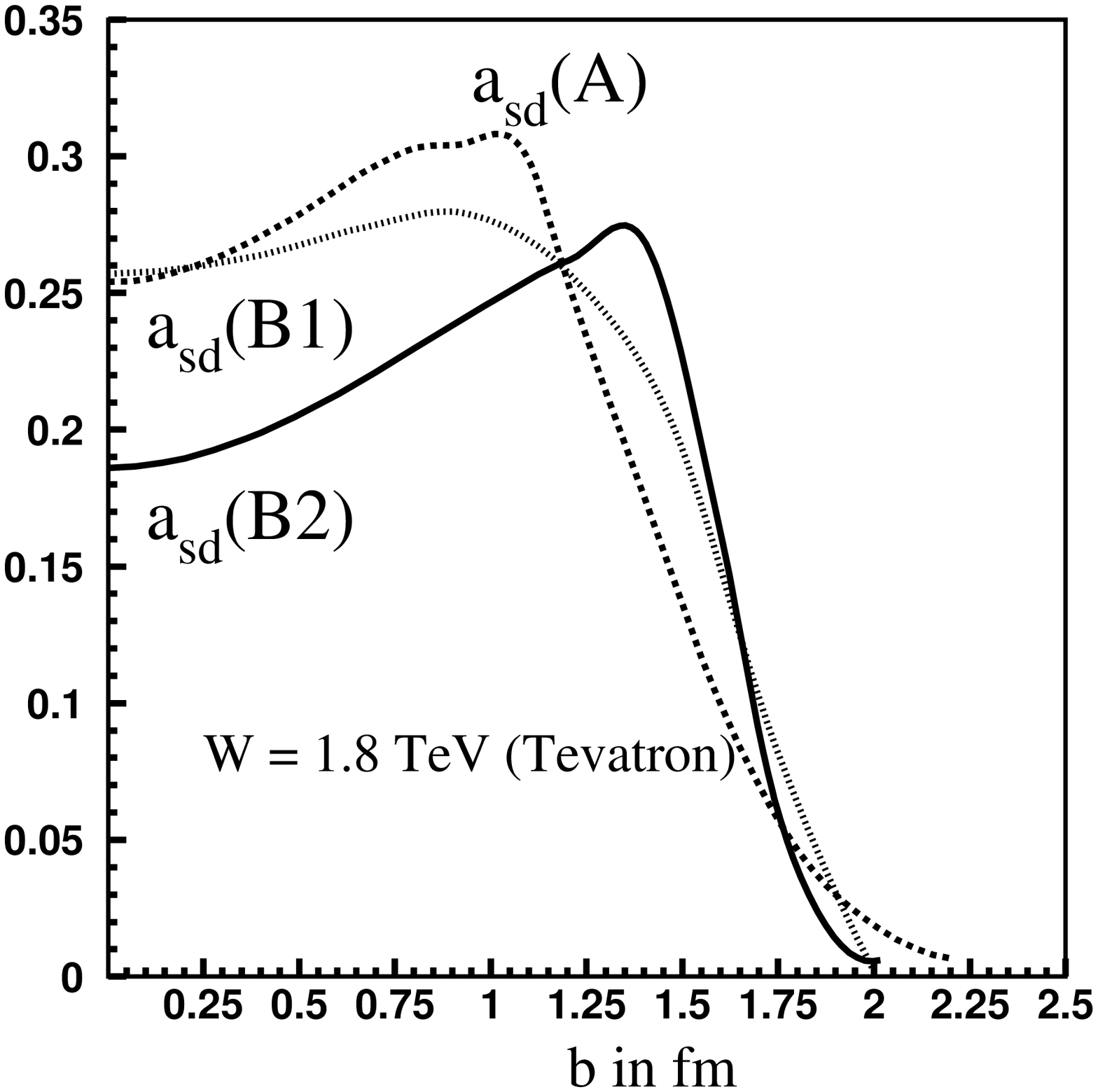,width=90mm,height=80mm} &
\epsfig{file=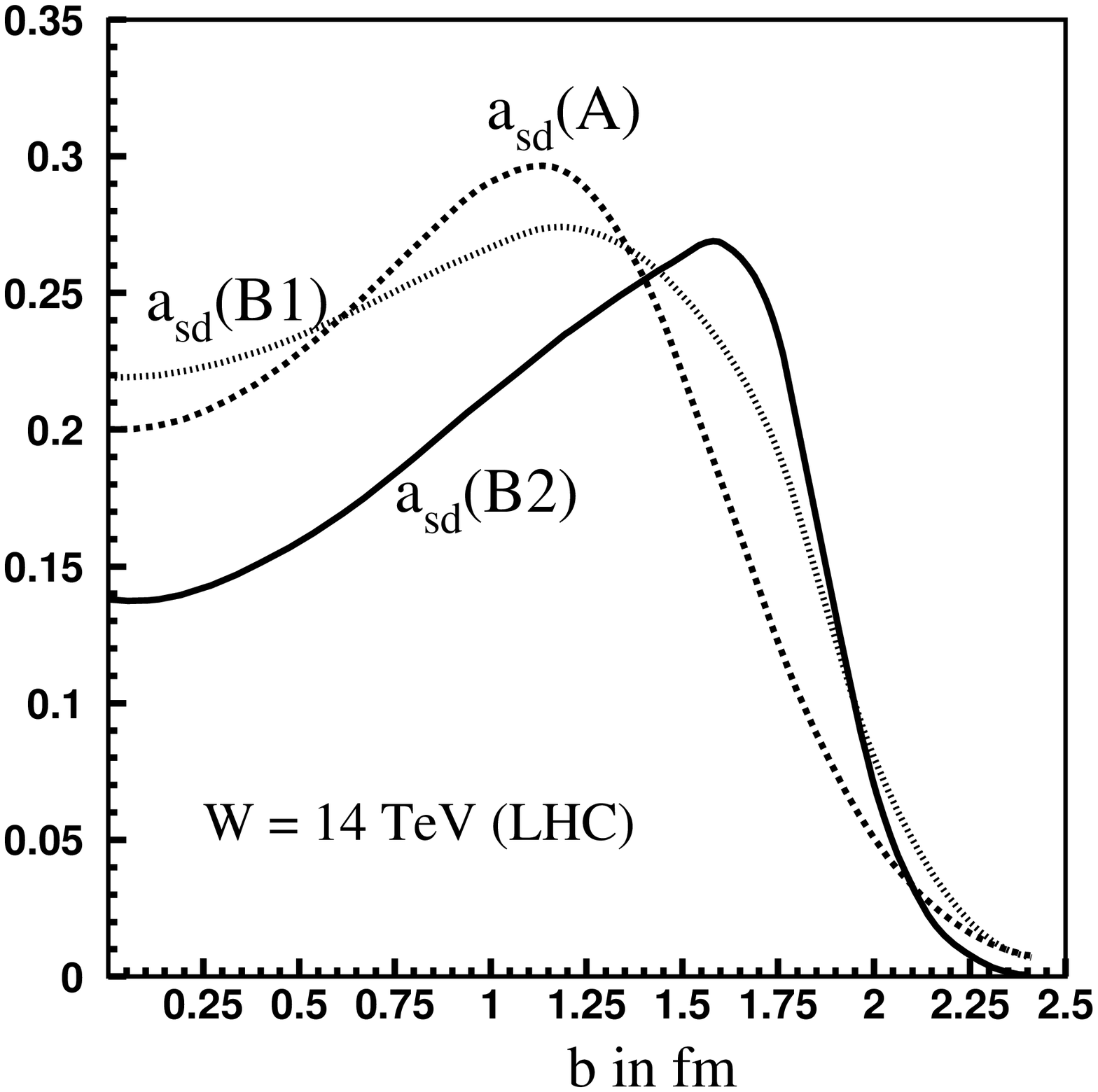,width=90mm,height=80mm} \\
\end{tabular}
\caption{Impact parameter dependence of $a_{sd}$
at the Tevatron and LHC in Models A, B(1) and B(2).}
\label{amsdbm}}

\par
The implication of the above is that $A_{2,2}$, which describes
the wave function scattering of $\Psi_2\times\Psi_2$,
has (even at W as low as a few $GeV$) a black central core in
$b$-space which expands with energy.
However, the wave function scattering, corresponding to $\Psi_1\times\Psi_1$ 
and $\Psi_1\times\Psi_2$,
are much smaller, well below the
black disc bound (see \fig{ampb}). Consequently, the amplitudes
$A_{1,1}$ and $A_{1,2}$ have a different behavior as functions of $(s,b)$,
(see \fig{ampb}).
\par
The behavior indicated above at the
Tevatron and LHC energies becomes more extreme at ultra high energies, when
$a_{el}$ reaches 1 at ever larger $b$ values (see Table 3).
When $a_{el}(s,b)$ = 1,
unitarity implies that at these $b$ values $a_{sd}\,=\,a_{dd}\,=\,0$. The
diffractive cross sections become, thus, highly peripheral and 
relatively small since they are confined 
exclusively to the very high $b$ values where the periphery of 
$a_{el}(s,b)$ is below the black disc bound. 
This prediction is observed in our calculation at ultra high energies, above the GZK 
ankle cutoff.
\DOUBLEFIGURE[ht]{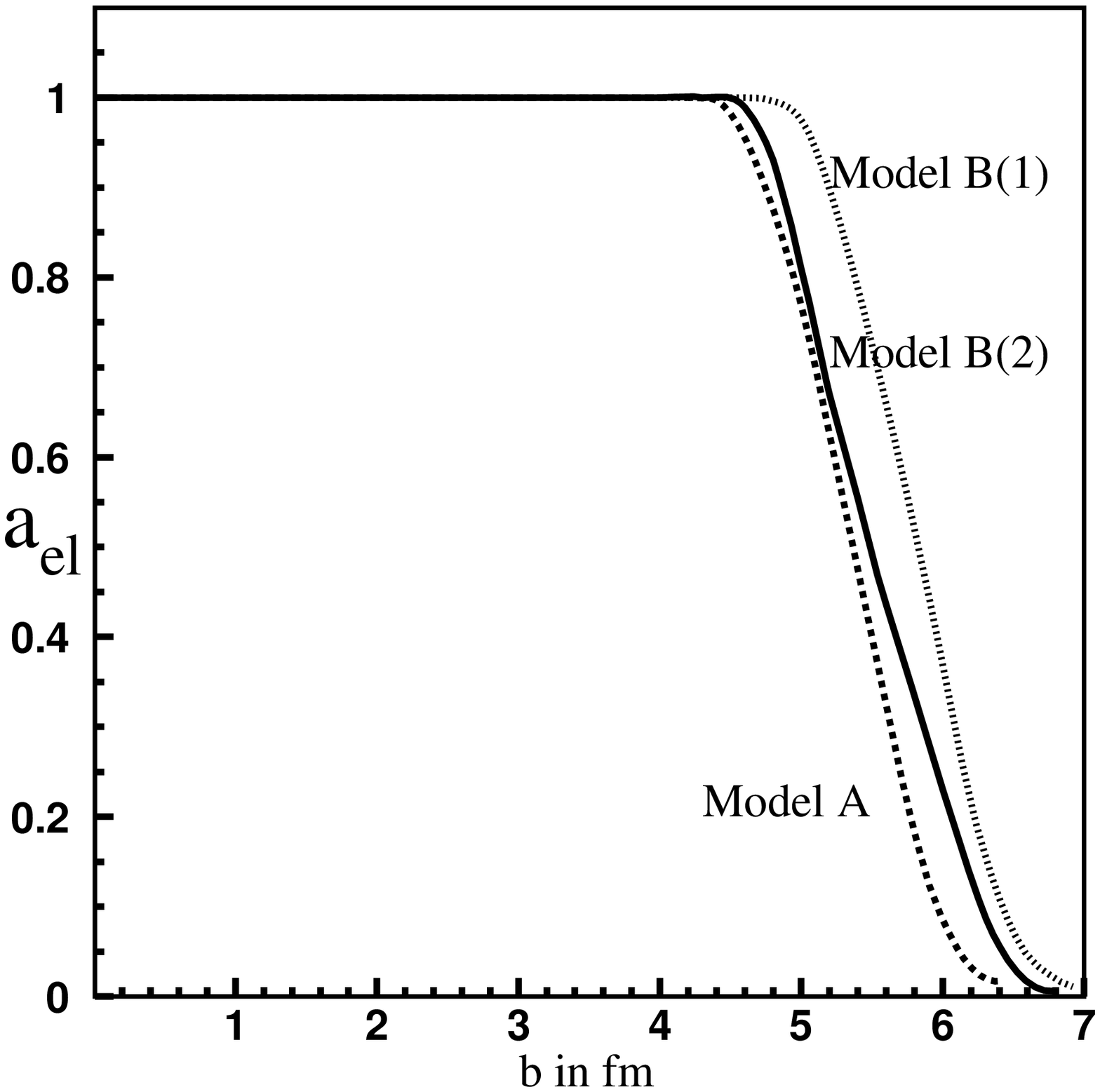,width=90mm,height=80mm}
{ 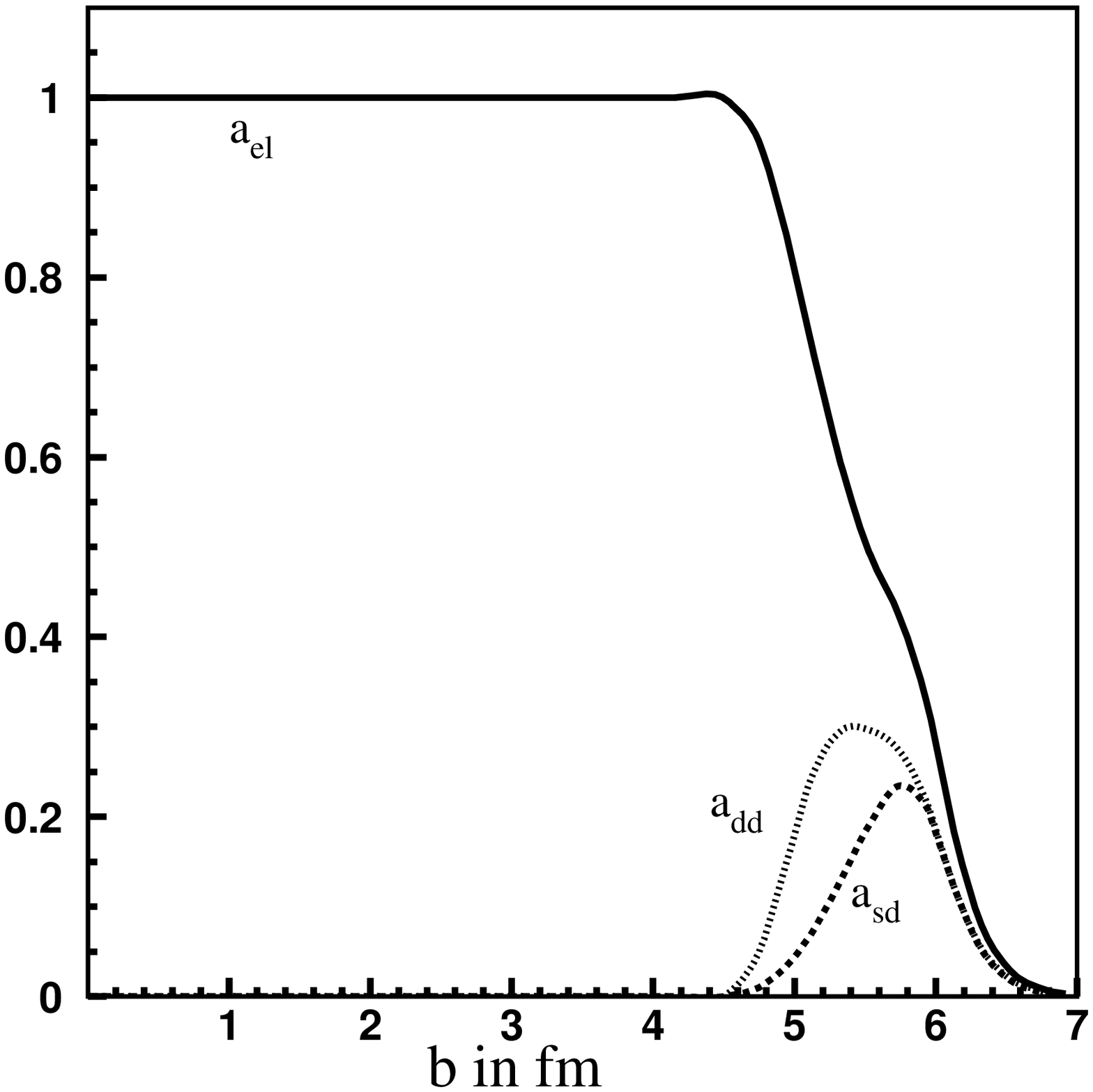,width=90mm,height=80mm}
{$b$ dependence, at the Planck mass, of $a_{el}$ in Models A, B(1) and 
B(2).
\label{pl2}}
{$b$ dependence of $a_{el}$, $a_{sd}$, $a_{dd}$ at the Planck mass
in Model B(2).
\label{cr107}}

\par
We demonstrate this feature and its consequence at the extreme Planck mass.  
\fig{pl2} clearly shows that as the black core of $a_{el}$ expands, the 
difference between Models A, B(1) and B(2), considered in this paper,  
diminishes, being 
confined to the narrow $b$ tail where $a_{el}(s,b)<1$.
\fig{cr107} compares 
$a_{el}$, $a_{sd}$ and $a_{dd}$ in Model B(2). Evidently, the diffractive 
amplitudes, which are logarithmically suppressed, survive just at the high $b$ 
tail of $a_{el}$. 
The above observations may be of interest in the analysis of Cosmic Ray 
experiments.

\section{Conclusions}
\par
In this paper we have presented a set of predictions for cross
sections, forward slopes and survival probabilities at ultra high energies 
focusing on the LHC.
There is a relatively small variance between the published predictions for
cross sections at the LHC. Our SD and DD slope predictions are novel and 
are
important for a proper assessment of the soft diffractive background to a
diffractively produced Higgs. 
\par
We wish to single out three significant properties of our model: 
\newline
1) We do not assume the existence of a soft Pomeron, and we obtain the best
fit with a parametrization which violates Regge (coupling constant) 
factorization.
This feature is in strong contradiction to the elementary features 
of a soft Pomeron which is defined as a pole in the complex J plane. 
\newline
2) The elastic amplitude is much smaller than unity in the accessible 
energy range.  
\newline
3) The survival probability for Higgs diffractive production at the LHC is very 
small, $S^2_{H}=0.7\%$.
\par
Our results for the cross sections and the slopes at Cosmic Ray energies up to
the GZK ankle are coupled to our analysis of the role of unitarity at high
energies. It may appear that our predictions contain 
two contradicting observations. On the one hand,
the centrality in $b$ space of the output elastic amplitude forces a 
peripheral $b$ space behavior on inelastic diffraction\cite{CTM}. This
feature  occurs already at  low energies, and is reflected in the
low values of the predicted survival probabilities, and the
consequent suppression of the soft and the hard inelastic diffraction.
On the other hand, we obtain a surprising result that the approach of the
elastic amplitude to the black disc bound in $b$ space, is so slow that
$a_{el}$ reaches unity at $b$=0 only at approximately W=30,000 $TeV$ in the 
c.m.s.
Hence, the elastic amplitude is below the black disc bound
all through the measurable Cosmic Ray range. Our result is in sharp
contrast to Frankfurt et al.\cite{S2FS} who claim that $a_{el}$ 
should reach the black bound already at the LHC.
\par
An intriguing question is how general our results are;
i.e. how reliable is our  simple parametrization of opacities, and our method
of imposing unitarity constraints?  The fact
that the parameters of our model depend on the parameterization of 
the Regge sector, introduces some, though small, level of uncertainty.
Another weakness is that the GLM model assumes
a Gaussian $b$ dependence of the input amplitudes.
This deficiency results in our inability to properly predict 
the diffractive dip structure and its position in $d \sigma_{el}/dt$. 
As we have seen our model reproduces well \cite{Rlet}, the
forward cone which contains more than 95\% of the data.
\par\FIGURE[ht]{\begin{minipage}{75mm}
\centerline{\epsfig{file=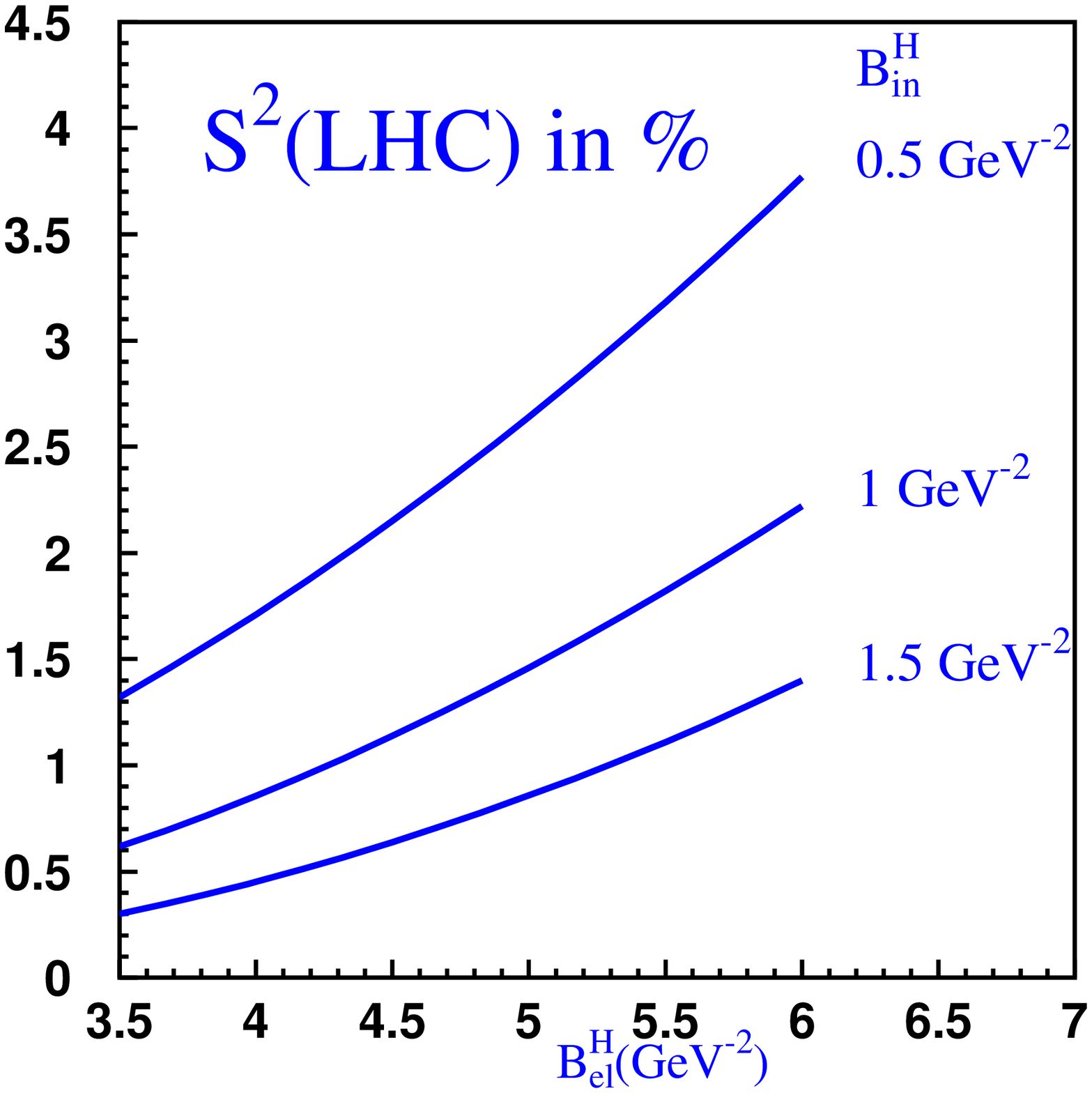,width=70mm}}
\end{minipage}
\caption{ The  dependence of $S^2$ at the LHC on $B^H_{el}$ and $ B^H_{in}$, 
the slopes for the hard cross sections.} 
\label{BH}}
Some additional clarifications concerning our three amplitude
Model B(2) prediction of $S_H^2=0.7\%$ are in place. This value is
significantly smaller than our two amplitude Model A
prediction of $S_H^2=2.7\%$.
We presume that our opacities at small $b$ do not introduce 
significant errors in the estimates of $S_H^2$. 
This has been explicitly shown in Ref.\cite{GKLMP} for the simple case of 
a single channel model, where we can assess the b-profile directly from the 
$\frac{d\sigma_{el}}{dt}$ data. For two channel models the $A_{i,k}$ 
amplitudes are model dependent and can not be determined directly from 
the data, hence the resulting $S_H^2$ is also model dependent.
In spite of this, we do have control over the output b-distribution of 
$a_{el}$ which can be compared with the data, 
very much like the single channel procedure.
This is the basis for our intuitive assessment that our estimate of $S_H^2$
is reasonably reliable, regardless of our choice for the b-profile. 
The results we obtain for $S_H^2$ depend on the values we choose for the 
elastic and inelastic hard slopes $B^H_{el}$ and $B^H_{in}$. We have 
determined these parameters from the HERA measured\cite{KOTE,PSISL} 
$J/\Psi$ photo and DIS production elastic and inelastic slopes. 
Our sensitivity to these parameters is shown in \fig{BH}. 
Note that when we change the value of $B_{in}^H$ we keep the ratio 
$V^2_{p \to d}/B^H_{in}$ unchanged. Doing so we do not change the 
cross section of the reaction 
$\gamma  + p \to J/\Psi + X \mbox{(M $\leq$ 1.6 GeV)}$. 
\par
A possible weak feature of our model is the fact that we possibly do not 
include Pomeron enhanced diagrams,  which in the Pomeron calculus are 
responsible for high mass diffraction.
The CDF collaboration \cite{CDFSD}fitting to a triple Pomeron formalism,
with $\alpha_{P}(0) = 1 + \Delta$,
 found that at the Tevatron energy:
 $\Delta = 0.112 \pm 0.010 \mbox{(stat)} \pm 0.011
\mbox{(syst)}$. 
We wish to draw the readers attention to the fact that in the triple 
Pomeron formalism \cite{Mueller}, the  diffractive mass 
distribution is divergent if $\Delta = 0$, and it converges for
 $\Delta > 0$.
 Hence, replacing the rich diffractive final states by the one state, 
we 
implicitly assumed that the mass spectrum of diffraction has a typical 
finite mass. 
Based on semi-hard processes  large mass diffraction has been
successfully
described in $e$-$p$ DIS\cite{DISMOD,KOTE}.
We hope to incorporate these processes in our model
as well. These semi-hard processes are potentially very important to the
calculation
of survival probabilities (see Refs.\cite{BBK,JM}) and can
decrease their value further\cite{JM}.
\par
Regardless of these deficiencies, we believe that a qualitative result similar
to ours will be obtained in any model in which the effect of unitarity is
significant.
These are open problems and should be further pursued. In particular,
since
we do not have a theory for non-perturbative QCD, the realibility of our
calculation should be checked by comparing it with the results obtained
for alternate models for soft interactions.

\section* {Acknowledgments}
We are grateful to Eran Naftali, Andrey Kormilitzin and Alex Prygarin for fruitful
discussions and technical help. We are indebted to Misha Ryskin with whom we had
a most enlightening and beneficial correspondence.
UM wishes to thank Daniele Treleani and Mark Strikman
for instructive discussions and correspondence.
We wish to sincerely thank our anonymous referee whose comments prompted us
 to improve  our presentation.
This research was supported
in part by the Israel Science Foundation, founded by the Israeli Academy of Science
and Humanities, by BSF grant $\#$ 20004019 and by
a grant from Israel Ministry of Science, Culture and Sport and
the Foundation for Basic Research of the Russian Federation.




\begin{thebibliography}{99}
\bibitem{Rlet}
M. Ryskin,
Private communication containing a calculation of
$\frac{d\sigma_{el}}{dt}$ in our Model B(2).
\end{thebibliography}
\end{document}